\documentclass[a4paper,11pt]{article}
\usepackage{jcappub} 

\arxivnumber{}
\title{\boldmath
Forecasting Coupled Dark Energy Parameters with the One-Loop Galaxy Power Spectrum}
\author[a]{Bilal T\"udes}

\author[b,c]{and Luca Amendola}%

\affiliation[a]{Institute of Theoretical Astrophysics,\\
Philosophenweg 12, 69120 Heidelberg, Germany
}

\affiliation[b]{%
 Institut f\"ur Theoretische Physik, Universit\"at Heidelberg,\\
Philosophenweg 16, 69120 Heidelberg, Germany
}%
\affiliation[c]{ New York University Abu Dhabi, PO Box 129188, Abu Dhabi, United Arab Emirates and Center for Astrophysics and Space Science (CASS), New York University Abu Dhabi}
\emailAdd{tuedes@thphys.uni-heidelberg.de}
\emailAdd{l.amendola@thphys.uni-heidelberg.de}

\date{\today}

\abstract{We forecast constraints on the parameters of the coupled dark energy model using DESI and Euclid data with the one-loop galaxy power spectrum. We investigate the distinguishability of our model from the zero-coupling scenario at the $1\sigma$ level and explore how the parameter constraints depend on the fiducial coupling $\beta_{\rm fid}$, the fixed potential slope parameter $\nu$,  the maximum wavenumber $k_{\max}$, and different prior choices.  We find that the inclusion of mildly non-linear scales improves the constraints on the coupling by roughly a factor of five. Then, we address the question of which is the minimum value of $\beta$ that can be distinguished from zero at $1\sigma$. We find that for our reference case  $k_{\rm max}=0.2 h/$Mpc,  $\beta=0.15$ lies roughly $1\sigma$ above zero. In the most optimistic case with $k_{\rm max}=0.3 h/$Mpc  and including a Planck prior on $\Omega_{m0}$, this value can be reduced to $0.05$. These values are substantially larger than the current constraints on $\beta$, but the latter  have been obtained assuming $\beta$ to be constant from at least the decoupling epoch to today, while we only employ late-time data. We conclude therefore that only models that allow for time-varying couplings can be detected with late-time datasets.}

\usepackage{graphicx}
\usepackage{comment}

\usepackage{amsmath}
\usepackage{dcolumn}
\usepackage{bm}
\usepackage{xcolor}
\usepackage{hyperref}
\renewcommand{\eqref}[1]{(\ref{#1})}

\usepackage{subcaption}
\usepackage[labelfont=bf, font=small, width=0.8\textwidth]{subcaption}

\begin{document}

\maketitle
\flushbottom
\section{\label{sec:int}Introduction}


The accelerated expansion of the Universe is well established by several independent observations \cite{Riess1998,Perlmutter1999,Eisenstein2005,Planck2020,Brout2022}. Although the $\Lambda$CDM model provides a successful description of these data, the physical nature of dark energy remains unknown \cite{PeeblesRatra2003,Copeland2006,Frieman2008}. This motivates studying models in which the dark sector has richer dynamics than a cosmological constant.

Coupled dark energy (CDE) provides one such possibility \cite{WangEtAl2024,Amendola2000,TocchiniValentiniAmendola2002,GumjudpaiEtAl2005,BoehmerEtAl2008,SinghSingh2016,BernardiLandim2017}. In these models, dark energy is described by a scalar field that interacts directly with cold dark matter. The interaction strength is controlled by the coupling parameter $\beta$.
The coupling changes both the background evolution and the growth of matter perturbations at linear and non-linear level. These modifications enter the redshift-space galaxy power spectrum through the linear growth factor $D(z)$ and growth rate $f(z)$ \cite{TudesAmendola2025,DiPortoAmendolaBranchini2012}.

In this work, we model the redshift-space galaxy power spectrum at one-loop order. The calculation includes the linear contribution, the $P_{13}$ and $P_{22}$ loop terms, EFT counterterms,   redshift-space distortion and damping, and shot noise \cite{Amendola_2022, PerkoEtAl2016,d_Amico_2020,Ivanov_2020}.  Extending the analysis to mildly nonlinear scales provides additional information, but requires nonlinear perturbation theory, galaxy bias, and a consistent treatment of short-scale effects \cite{Bernardeau2002,Carrasco2012,PerkoEtAl2016,Desjacques2018}. We solve for the CDE background, linear growth, and time-dependent perturbation kernels instead of assuming the usual Einstein--de Sitter (EdS) kernels. Alcock--Paczyński distortions \cite{AlcockPaczynski1979,Lepori_2017} are also included  in the observed spectrum. To increase the level of model-independence, we refrain from modeling the linear power spectrum shape based on specific inflationary models, and we choose instead to leave $P_0(k)$ free in a number of wavebands.

We use the Fisher-matrix formalism \cite{Amendola_2022,Tegmark1997,SeoEisenstein2003} to forecast constraints from DESI \cite{desicollaboration2016,desicollaboration2} and Euclid\ \cite{Euclid2025}. Our main target is the coupling $\beta$, which is varied jointly with $\Omega_{m0}$ for fixed choices of the potential slope $\nu$. The analysis varies the binned linear spectrum $ P_0(k)$ together, in each redshift bin, with the bias, counterterm, damping, and shot-noise parameters. 

   This work has two central goals. First, we aim at ascertaining how non-linear scales affect the power of the experiments in constraining the main parameters, in particular the coupling $\beta$. Secondly, we aim at finding the lowest value of $\beta$ that can still be distinguished from zero, i.e. from an uncoupled model. In doing this, we allow exploration of values of $\beta$ much larger than those that have already been constrained by present observations, which is around $\beta\approx 0.02$ (see e.g., \cite{Chakraborty:2025syu,gómezvalent2026,Amendola_Barros_2019}). The reason is that, so far, most constraints have been obtained for a constant coupling throughout the entire cosmic evolution, and consequently, they have been found to be quite restrictive. We expect that a varying $\beta$ can therefore considerably loosen the current observational bounds.

\section{The coupled dark energy model}

We give here a quick summary of the CDE model; for further details, see Refs.~\cite{TudesAmendola2025,Amendola2000,Amendola2004}. We consider a spatially flat FLRW universe in which the dark-energy scalar field (subscript $\phi$) interacts with the nonrelativistic matter sector (subscript $m$). This interaction transfers energy and momentum between the two components. Therefore, their energy--momentum tensors are not conserved separately and satisfy:

\begin{equation}
\nabla_\mu T_{\nu (\phi)}^\mu=\beta T_m \nabla_\nu \phi \, , \quad 
\quad \nabla_\mu T_{\nu(m)}^\mu=-\beta T_m \nabla_\nu \phi \, .\label{eq:energy-mom}
\end{equation}

Here, $\beta$ is a constant dimensionless coupling that controls the interaction strength between the dark-energy and dark-matter sectors, while $T_m$ denotes the trace of the energy--momentum tensor of pressureless matter. Baryons are assumed to remain uncoupled, ensuring consistency with local tests of gravity. We also neglect radiation, since our analysis is restricted to late-time cosmological evolution. Although in principle $\beta$ can vary arbitrarily through the cosmic evolution, we assume here it to be a constant, since our dataset is limited to a relatively small range of redshifts.

Taking the time component of Eq.~\eqref{eq:energy-mom} yields the background continuity equations for the dark-matter and scalar-field energy densities, respectively:

\begin{align}
    \rho_{m}' +3(P_{m}+ \rho_{m})&=-\beta\rho_{m}\phi'
    \label{eq:matterdens},
\\  
 \rho_{\phi}' +3(P_{\phi}+ \rho_{\phi})&=\beta\rho_{m}\phi'
\label{eq:scalar_field_dens}.
\end{align}

Hereafter, a prime denotes differentiation with respect to $\eta \equiv \ln a$, where $a$ is the scale factor.

The Friedmann equation is given by:

\begin{equation}
    3H^{2}= \rho_{\phi}+\rho_{m}\, ,
    \label{eq:H/H'}
\end{equation}
where we adopted units such that $8\pi G=c=1$.
The evolution of the scalar field is described by the coupled Klein--Gordon equation,

\begin{equation}
\label{eq:KG}
    \phi''+(3+\frac{H'}{H})\phi'+\frac{1}{H^2}\frac{dV}{d\phi}=3\beta \Omega_m \, .
\end{equation}
The background equations can be written in dimensionless form by defining

\begin{equation}
x=\frac{\phi'}{\sqrt{6}},\quad y=\frac{1}{H}\sqrt{\frac{V}{3}}\, ,
\label{eq:x,y}
\end{equation}
which gives
\begin{eqnarray}
x' && =\sqrt{\frac{3}{2}}\beta\left(1-x^{2}-y^{2}\right) +  \sqrt{\frac{3}{2}}\nu y^{2}+\frac{1}{2}x\left(3x^{2}-3y^{2}-3\right) \, ,
\label{eq:x}\\
y' && =-\sqrt{\frac{3}{2}}\nu xy-\frac{1}{2}y\left(-3x^{2}+3y^{2}-3\right)  \, ,
\label{eq:y}\\
H' &&=-\frac{1}{2}H\left(3x^{2}-3y^{2}+3\right)  \, . \label{eq:h}  
\end{eqnarray}

We adopt an exponential scalar-field potential, $V(\phi)=V_0 e^{-\nu\phi}$, where $\nu $ is the potential-slope constant parameter. The matter and scalar-field density parameters are defined as $\Omega_m = 1-x^2-y^2$ and $\Omega_\phi = x^2+y^2$, respectively (with present-day fiducial values $\Omega_{m0}=0.31$ and $\Omega_{\phi0}=0.69$). The background evolution governed by Eqs.~\eqref{eq:x}--\eqref{eq:h} has been studied extensively in the literature~\cite{Amendola2000,GumjudpaiEtAl2005,BoehmerEtAl2008,BahamondeEtAl2018}. To solve the system, we specify initial conditions at the phase-space critical point corresponding to the $\phi$-matter-dominated epoch ($\phi$MDE).  This is a natural starting point because the CDE trajectory passes through the $\phi$MDE regardless of the potential's detailed form.

At linear order, the evolution of matter perturbations is described by the growth factor $D$, which satisfies the following equation:

\begin{eqnarray}
    &&D''+FD'-SD=0\,.
    \label{eq:D}
\end{eqnarray}
Defining the growth rate as $f \equiv D'/D$, Eq.~\eqref{eq:D} becomes:

\begin{equation}
    f'+f^2+Ff-S=0\,. \label{eq:f}
\end{equation}
In the CDE model, the background evolution described by $x$ and $y$ enters Eqs.~\eqref{eq:D} and~\eqref{eq:f} through the friction function $F$ and source term $S$, defined as:

\begin{eqnarray}
    F&=&\frac{1}{2}-\frac{3}{2}(x^2-y^2)-\sqrt{6}\beta x  \, , \label{eq:F}\\
     S&=&\frac{3}{2}[(1+2\beta^2)(1-x^2-y^2)]  \, .
    \label{eq:S}
\end{eqnarray}

The  background functions entering our power-spectrum calculation are the growth factor $D$ and growth rate $f$, which depend on $\beta$, $\nu$, and the initial values of $x$ and $y$. These initial conditions are chosen so that the cosmological evolution passes through the $\phi$MDE and reproduces the adopted present-day value of $\Omega_{m0}$; see Ref.~\cite{TudesAmendola2025} for further details. In conclusion, background and linear perturbation growth depend only on $\beta,\nu$ and $\Omega_{m0}$.

\section{\label{sec:Pw}One-loop power spectrum}

In this section we write the  galaxy power spectrum as a sum of a linear contribution, the leading nonlinear (1-loop) correction, UV counterterms   and shot noise \cite{Ivanov_2020, d_Amico_2020,Amendola_2022}. 

\begin{equation}
    P_{gg}(k, \mu, z)=S_{\mathrm{g}}(k, \mu, z)^2\left[P^{\rm lin}(k, \mu, z)+P^{\rm 1\text{-}loop}(k, \mu, z)+P^{\rm UV}(k, \mu, z)\right]+P^{\rm sn}(z) ,
\end{equation}
where $k=\sqrt{k_{\parallel}^2+k_{\perp}^2}$ and $\mu\equiv k_{\parallel}/k$, with $k_{\parallel}$ ($k_{\perp}$) denoting the component of the wavevector parallel (perpendicular) to the line of sight. We implement the Alcock--Paczyński mapping between fiducial and model coordinates following the definitions of \cite{Amendola_2022}. The galaxy power spectrum  $P_{gg}(k,\mu,z)$ depends on the linear power spectrum and on a set of nuisance parameters describing galaxy bias and short-scale physics. In particular, we include four bias parameters, $\bigl(b_1(z),\, b_2(z),\, b_{G_2}(z),\, b_{\Gamma_3}(z)\bigr)$, which relate the galaxy density field to the underlying matter distribution, and three UV counterterms, $\bigl(c_0(z),\, c_2(z),\, \tilde c(z)\bigr)$, which capture the impact of unresolved small-scale physics within effective field theory.

We write $S_{\mathrm{g}}(k,\mu,z)$ as a smoothing term that accounts for both spectroscopic redshift errors and the Finger-of-God (FoG) effect \cite{Euclid4,Beutler_2016}:
\begin{equation}
    S_{\mathrm{g}}(k,\mu,z)
=\exp\!\left[-\frac{1}{2}\,(k\mu\sigma_z)^2\right]\,
 \exp\!\left[-\frac{1}{2}\,(k\mu\sigma_f)^2\right]\,,
\end{equation}
where
\begin{equation}
    \sigma_z=\sigma_0(1+z)\,H(z)^{-1}\,.
\end{equation}

The linear contribution is given by the standard expression,
\begin{equation}
    P^{\rm lin}(k,\mu,z)=\left(b_1+\mu^2\,f(z)\right)^2\,P(k,z)\,,
\end{equation}
where $b_1$ is the linear bias parameter and $f$ is the scale-independent linear growth rate. The linear real-space matter power spectrum is

\begin{equation}
    P(k, z)= \Big(\frac{D(z)}{D(0)}\Big)^2P_0(k) \,.
\end{equation}
We obtain the present-day linear spectrum  $P_0(k)$ from the CLASS code  \cite{Diego_Blas_2011} and evolve it using the linear growth function $D$. 

The term $P^{\rm 1\text{-}loop}(k,\mu,z)$ denotes the leading nonlinear correction to the linear order, arising from mode coupling in perturbation theory (schematically the sum of the  $P_{22}$ and $P_{13}$ contributions,  in redshift  galaxy space). The complete expressions for $P^{\rm 1\text{-}loop}(k,\mu,z)$ and $P^{\rm UV}(k, \mu, z)$ are given in the  appendix of \cite{Amendola_2022}.

The shot-noise contribution is modeled as:
\begin{equation}
P^{\rm sn}(z)=\frac{1}{n(z)}\left(1+P_{\mathrm{shot}}\right)\,,
\end{equation}
where $n(z)$ is the  number density of galaxies and $P_{\mathrm{shot}}$ parametrizes  deviations from pure Poisson shot noise.

We adopt the CDE background and perturbation formalism of \cite{TudesAmendola2025}, including the initial conditions associated with the $\phi$MDE phase. Within this framework, we solve for the growth function $D$ and growth rate $f$ for each value of $\beta,\Omega_{m0},\nu$  and use the full time-dependent CDE perturbation kernels rather than EdS-like kernels. The resulting spectrum depends on the  coupling $\beta$ and the potential-slope parameter $\nu$. 

\begin{figure}[htbp]
    \centering
    \begin{minipage}[t]{0.49\linewidth}
        \centering
        \includegraphics[width=\linewidth]{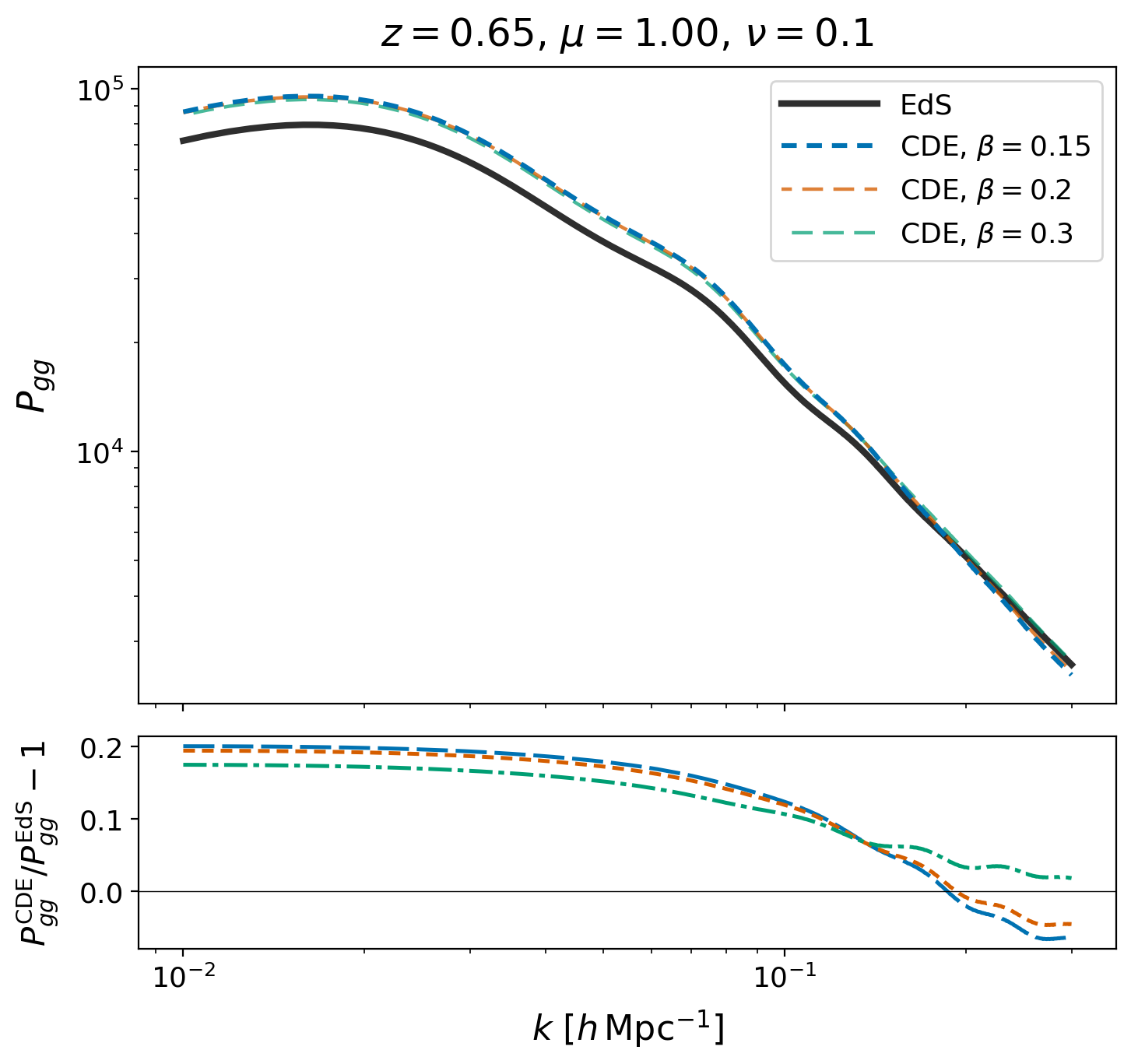}
        \textbf{(a)} $z=0.65$: CDE compared with EdS
    \end{minipage}
    \hfill
    \begin{minipage}[t]{0.49\linewidth}
        \centering
        \includegraphics[width=\linewidth]{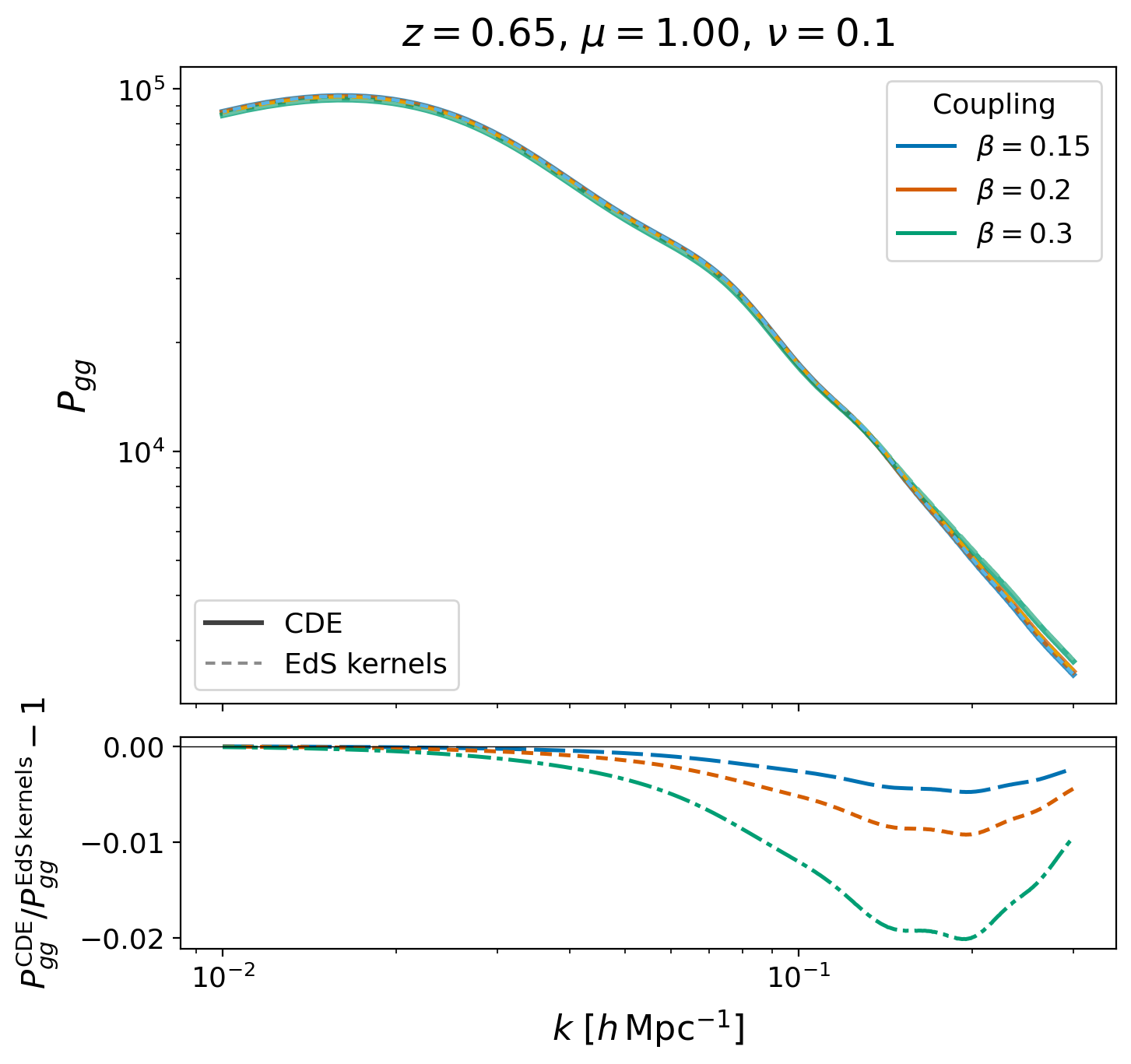}
        \textbf{(b)} $z=0.65$: CDE compared with EdS kernels
    \end{minipage}

    \par\medskip
    \begin{minipage}[t]{0.49\linewidth}
        \centering
        \includegraphics[width=\linewidth]{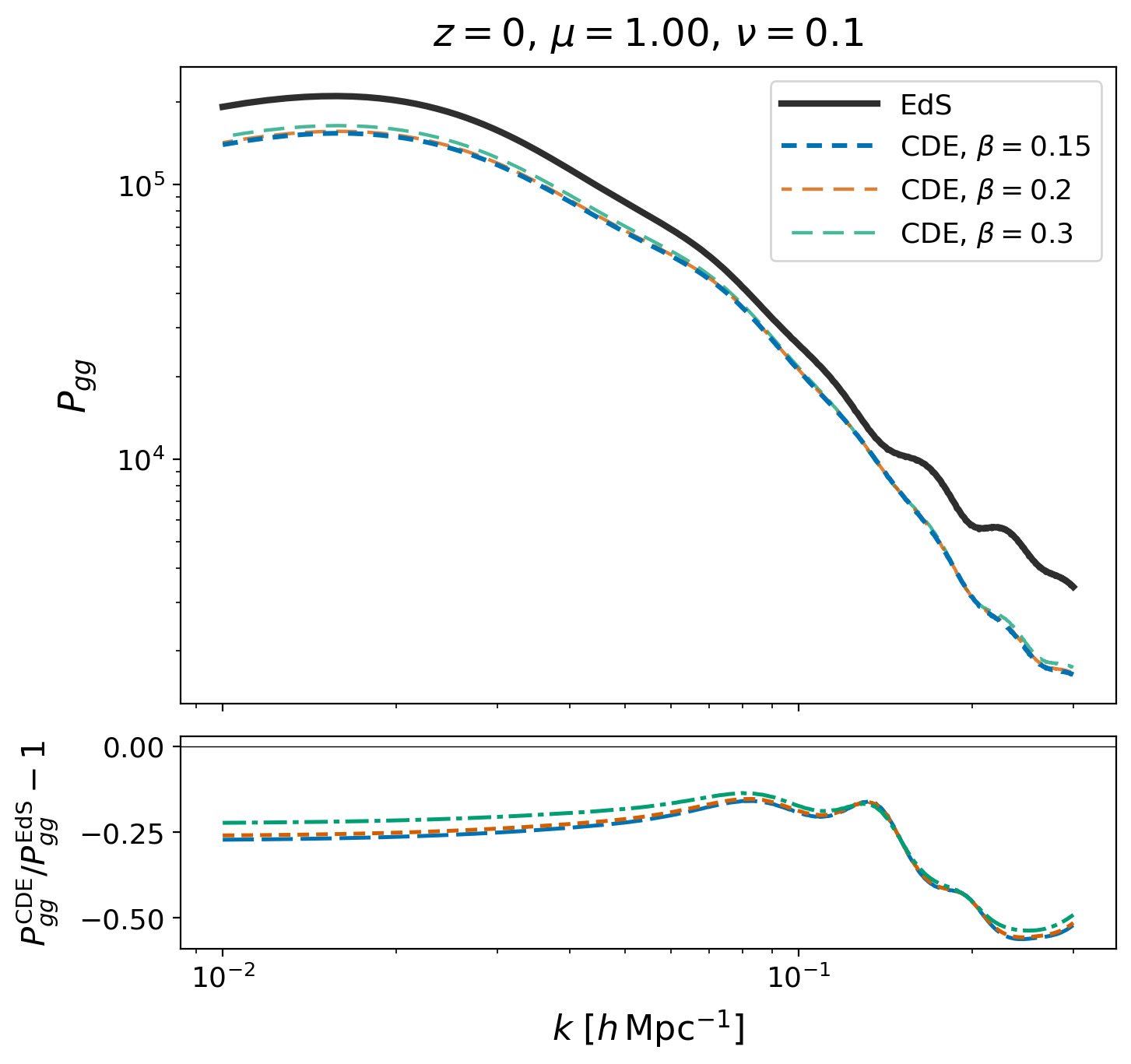}
        \textbf{(c)} $z=0$: CDE compared with EdS
    \end{minipage}
    \hfill
    \begin{minipage}[t]{0.49\linewidth}
        \centering
        \includegraphics[width=\linewidth]{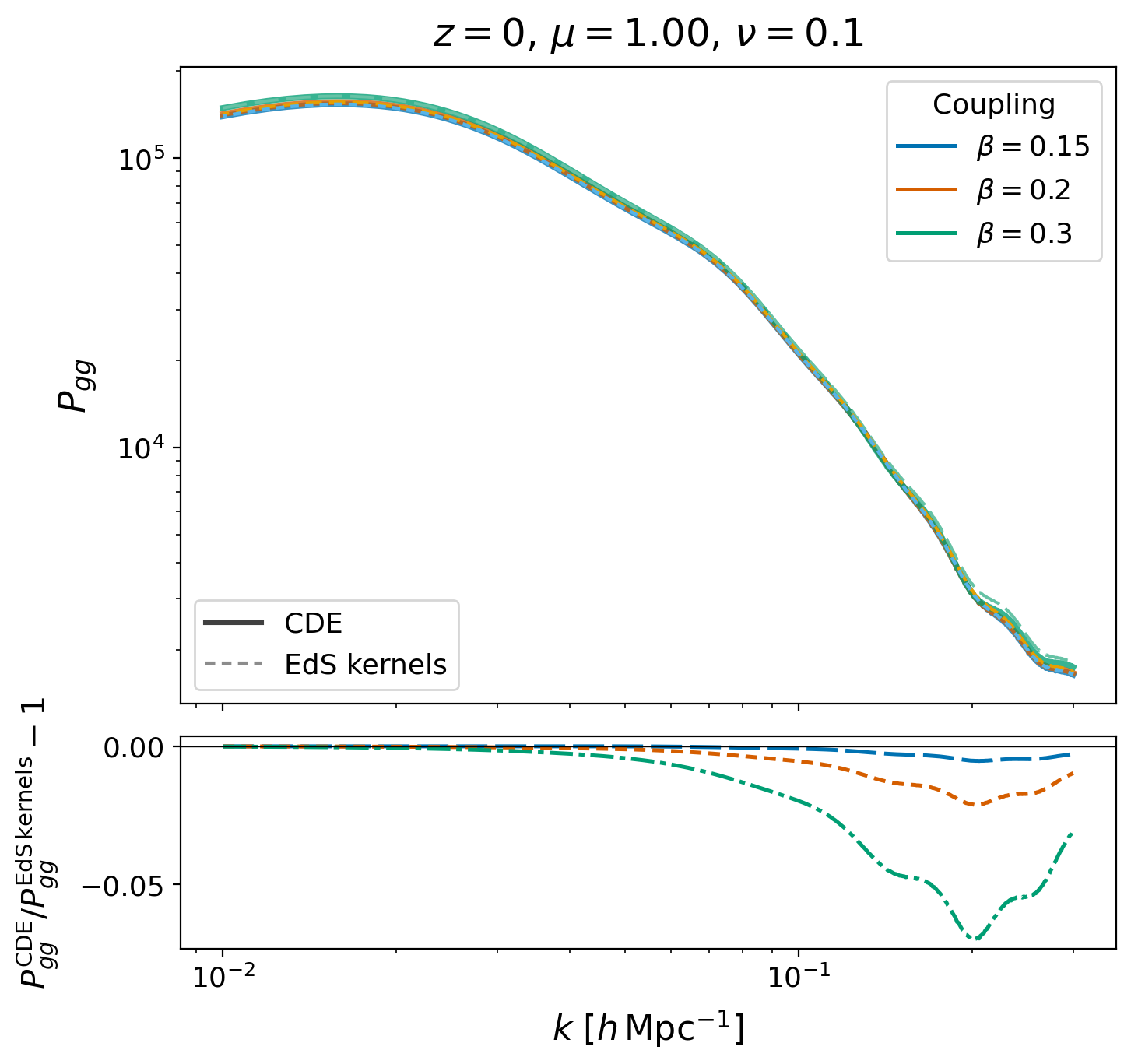}
        \textbf{(d)} $z=0$: CDE compared with EdS kernels
    \end{minipage}
    \caption{Comparison of the galaxy power spectra and their relative differences at $z=0.65$ (top row) and $z=0$ (bottom row).
    Left panels compare CDE with the full EdS model, while the right panels compare CDE with
    the EdS-kernel approximation. All panels use $\mu=1$, $\nu=0.1$, and $\beta=0.15$, $0.2$, and $0.3$.}
    \label{fig:pg_comparisons}
\end{figure}

To quantify how much the CDE power spectrum deviates from the EdS   model, Figure~\ref{fig:pg_comparisons} presents two comparisons. First, we compare the CDE spectrum with a full EdS model, using both the EdS background and EdS perturbation kernels. Second, we compare it with a model that retains the CDE background but replaces the CDE kernels with EdS kernels to isolate the effect of the kernels on the power spectrum.  The large difference in the first comparison shows that the deviation is mainly caused by the modified CDE background and growth functions. The much smaller difference in the second comparison indicates that the time dependence of the CDE kernel coefficients has only a minor effect at $z=0.65$. At $z=0$, however, the difference reaches about $7\%$ for $\beta=0.3$, so in general it cannot be neglected. For a comparison of EdS and time-dependent kernels in a different interacting dark-sector model, see Ref.~\cite{wolney}.

\section{\label{sec:Fs}Fisher Matrix}

To forecast parameter constraints, we adopt the Fisher-matrix formalism for the redshift-space galaxy power spectrum \cite{Tegmark1997,Amendola_2022,Euclid4}. Assuming Gaussian statistics for the Fourier modes, the Fisher matrix provides an estimate of the expected parameter covariance around a fiducial model and can be written as a sum over redshift bins and wavenumber shells.

For a given redshift bin, the Fisher matrix can be expressed as:

\begin{equation}
F_{\alpha\beta}
=\frac{1}{8\pi^{2}}\int_{-1}^{+1}\! d\mu
\int_{k_{\min}}^{k_{\max}}\! k^{2}\,dk\,
\frac{\partial \ln P_{gg}(k,\mu,z)}{\partial \theta_{\alpha}}\,
\frac{\partial \ln P_{gg}(k,\mu,z)}{\partial \theta_{\beta}}V\, ,
\end{equation}
where $V$ is the survey volume and $\theta_{\alpha}$ denotes a parameter of the model. We defined 20 $k$-bins as  [ $k_1,...,k_{20}$]. The bins are uniformly spaced in $k^2$, starting from $k_{\min}=0.001\,h\,\mathrm{Mpc}^{-1}$, for the upper limits $k_{\max}=0.20$, $0.25$, and $0.30\,h\,\mathrm{Mpc}^{-1}$. We also consider $k_{\max}=0.10\,h\,\mathrm{Mpc}^{-1}$ as a linear-scale reference scale.
 The model parameters $\theta_\alpha$ used in the  forecast are organized as
 \begin{equation}
\boldsymbol{\theta}
=
\left[
\boldsymbol{g},
\boldsymbol{p},
\boldsymbol{n}
\right],
\label{eq:parspace}
\end{equation}
\begin{equation}
\boldsymbol{g}=(\beta,\Omega_{m0}),
\quad 
\boldsymbol{p}
=
\left(
\ln P_0(k_1),\ldots,\ln P_0(k_{N_k})
\right),
\quad
\boldsymbol{n}=
(\boldsymbol{n}_1,\ldots,\boldsymbol{n}_{N_z}).
\label{eq:parspace2}
\end{equation}
We  divide the parameters into three groups: the global  parameters $\boldsymbol{g}$, the binned linear-spectrum  $\boldsymbol{p}$, and the nuisance parameters $\boldsymbol{n}$.  $ P_0(k_j)$ denotes the linear power spectrum
in the $j$th wavenumber bin. Within the nuisance group, $\boldsymbol{n}_i$ collects the parameters associated with redshift bin $i$:
\begin{equation}
\boldsymbol{n}_i
=
\left(
\ln b_{1,i},
b_{2,i},
b_{G_2,i},
b_{\Gamma_3,i},
c_{0,i},
c_{2,i},
\tilde c_i,
\ln\sigma_{f,i},
P_{\mathrm{shot},i}
\right).
\end{equation}
  The Fisher matrix has the following block structure

\begin{equation}
F=
\begin{pmatrix}
\displaystyle\sum_i F_{gg}(z_i)
&
\displaystyle\sum_i F_{gp}(z_i)
&
F_{g n_1}
&
F_{g n_2}
&
\cdots
\\[6pt]
\displaystyle\sum_i F_{pg}(z_i)
&
\displaystyle\sum_i F_{pp}(z_i)
&
F_{p n_1}
&
F_{p n_2}
&
\cdots
\\
F_{n_1 g}
&
F_{n_1 p}
&
F_{n_1 n_1}
&
0
&
\cdots
\\
F_{n_2 g}
&
F_{n_2 p}
&
0
&
F_{n_2 n_2}
&
\cdots
\\
\vdots
&
\vdots
&
\vdots
&
\vdots
&
\ddots
\end{pmatrix}.
\label{eq:mat1}
\end{equation}

\begin{itemize}
    \item $F_{gg}(z_i)$ is the $2\times2$ global-parameter block
    at redshift bin $z_i$.

    \item $F_{pp}(z_i)$ is the $N_k\times N_k$ linear-power-spectrum band block
    at redshift bin $z_i$.

    \item $F_{gp}(z_i)$ and $F_{pg}(z_i)=F_{gp}^{\mathrm T}(z_i)$ are the $2\times N_k$ and $N_k\times2$ blocks coupling the global
    parameters to the linear-spectrum bands.

    \item $F_{g n_i}$ and $F_{n_i g}=F_{g n_i}^{\mathrm T}$ are the $2\times N_{n_i}$ and $N_{n_i}\times2$ blocks coupling the global
    parameters to the nuisance parameters of redshift bin $i$.

    \item $F_{p n_i}$ and $F_{n_i p}=F_{p n_i}^{\mathrm T}$ are the $N_k\times N_{n_i}$ and $N_{n_i}\times N_k$ blocks coupling the
    linear-spectrum bands to the nuisance parameters of redshift bin $i$.

    \item $F_{n_i n_i}$ is the $N_{n_i}\times N_{n_i}$ nuisance-parameter block of redshift
    bin $i$, containing the bias, counterterm, damping, and shot-noise
    parameters.
\end{itemize}

Here, $N_k$ denotes the number of $\ln P_0(k)$ bands and $N_{n_i}$ the number of nuisance parameters in the redshift bin $i$. The $gg$, $gp$, $pg$, and $pp$ blocks are summed over all redshift bins, whereas the blocks involving $\boldsymbol{n}_i$ remain redshift bin dependent. For the linear power spectrum, each mode $k_i$ is independent, and taking derivatives with respect to the $ P_{0}(k_i)$ bands gives a diagonal $F_{pp}$. In the one-loop case, however, the non-linear contributions $P_{22}$ and $P_{13}$ mix  different modes, generating off-diagonal terms.

\section{\label{sec:Sr}Survey setup}
We perform our forecasts using two  galaxy surveys, Euclid  ($z \sim 0.9$--$1.8$) \cite{Euclid2025} and DESI ($z \sim 0.6$--$1.3$) \cite{desicollaboration2016,desicollaboration2}, which probe different galaxy populations and redshift ranges. The number densities and volumes for each redshift bin of DESI and Euclid used in our forecasts are presented in Tables~\ref{tab:desi} and~\ref{tab:euclid}, respectively.

\begin{table}[ht]
\centering

\begin{tabular}{|c |c |c |}
\hline
$z$ 
& $V\,(h^{-1}\mathrm{Gpc})^3$ 
& $10^{3}\cdot n\,(h\,\mathrm{Mpc}^{-1})^{3}$ 
\\
\hline
0.6--0.7 & 2.43 & 0.657 \\\hline
0.7--0.8 & 2.89 & 1.58  \\\hline
0.8--0.9 & 3.31 & 1.09  \\\hline
0.9--1.0 & 3.69 & 0.897 \\\hline
1.0--1.1 & 4.03 & 0.518 \\\hline
1.1--1.2 & 4.32 & 0.444 \\\hline
1.2--1.3 & 4.57 & 0.410 \\
\hline
\end{tabular}
\caption{DESI (LRG+ELG)}
\label{tab:desi}
\end{table}

\begin{table}[htbp]
\centering

\begin{tabular}{|c |c |c |}
\hline
$z$ 
& $V\,(h^{-1}\mathrm{Gpc})^3$ 
& $10^{3}\cdot n\,(h\,\mathrm{Mpc}^{-1})^{3}$ 
\\
\hline
0.9--1.1 & 7.94 & 0.686 \\\hline
1.1--1.3 & 9.15 & 0.558 \\\hline
1.3--1.5 & 10.1 & 0.421 \\\hline
1.5--1.8 & 16.2 & 0.261 \\
\hline
\end{tabular}
\caption{Euclid}
\label{tab:euclid}
\end{table}
The fiducial parameter values adopted in this work are identical to those discussed and used in \cite{Amendola_2022},   and are reported in Table~\ref{tab:fiducial}. These fiducial values are adopted from the BOSS analysis of Ref.~\cite{Ivanov_2020}, except for $b_1$, which is chosen separately for DESI and Euclid, and $b_{\Gamma_3}$, which is set to zero.

\begin{table}[ht]
    \centering
   
    \resizebox{\textwidth}{!}{%
    \begin{tabular}{|c |c |c |c |c |c |c |c |c |c |c |c |c |}
        \hline
        $b_1^{\rm DESI}$ 
        & $b_1^{\rm Euclid}$ 
        & $b_2$ 
        & $b_{G_2}$ 
        & $b_{\Gamma_3}$ 
        & $c_0$ 
        & $c_2$ 
        & $\tilde c$& $P_{\rm shot}$ 
        & $\sigma_f$ 
        & $\sigma_0$ & $\beta$ & $\Omega_{m0}$ \\
        \hline
        1.99 & 1.65 & $-3.21$ & 0.545 & 0.0 & $-53.0$ & $-21.0$ & 187.0 & 0& 5.0 & 0.0 & 0.15& 0.31  \\
        \hline
    \end{tabular}%
    }
     \caption{Fiducial parameters. The linear bias differs between DESI and Euclid. The counterterms $c_0$ and $c_2$ are given in $(h^{-1}\mathrm{Mpc})^2$, $\tilde c$ in $(h^{-1}\mathrm{Mpc})^4$, and $\sigma_f$ in $h^{-1}\mathrm{Mpc}$; $P_{\rm shot}$, $\beta$, and $\Omega_{m0}$ are dimensionless. }
    \label{tab:fiducial}
\end{table}
For the fiducial linear matter power spectrum $P_0(k)$, we adopt $\Lambda$CDM with $h=0.67$, $\Omega_b=0.049$, $\Omega_{\rm cdm}=0.265$, $\sigma_8=0.81$, and $n_s=0.965$, with $\Omega_k=0$.  In addition, CDE  introduces two extra parameters, $\beta$ and $\nu$. Following most literature, we take as reference value $k_{\rm max}=0.2 h/$Mpc, but we show results also for the linear regime $k_{\rm max}=0.1 h/$Mpc and for an "aggressive" non-linear one, $k_{\rm max}=0.3 h/$Mpc. We explore a range of fiducial values of $\beta_{\rm fid}$ to determine when it can be distinguished from zero at the $1\sigma$ level. For illustration, most figures use $\beta_{\rm fid}=0.15$, the lowest sampled value distinguishable from zero in the joint DESI+Euclid forecast at $\nu=0.1$ and   $k_{\max}=0.2\,h\,{\rm Mpc}^{-1}$. Since the parameters $\beta$ and $\nu$ are strongly correlated, we fix the fiducial value of $\nu$ to a set of representative values, namely $\nu= 0,\; 0.05,\; 0.1,\; 0.15,\; 0.2$.

\section{\label{sec:Rs}Results}
\subsection{Impact of priors on $\beta$ constraints}

In our approach, the linear power spectrum shape is free to vary in several wavebands. This avoids restricting the initial conditions to any specific inflationary model, but on the other hand, it might allow for unphysical wild oscillations. It is then worth investigating the effect of some  smoothing over the spectrum shape. This can be done by adding an off-diagonal prior on the  entries of the spectrum Fisher matrix. We model the smoothing prior as

\begin{equation}
\left(\Pi_{\rho}\right)_{ij}
=
\begin{cases}
0, & i=j,\\[6pt]
\rho\,
\exp\!\left[-\left(\dfrac{|k_i-k_j|}{k_{\rm corr}}\right)^2\right]
\sqrt{(F_{pp})_{ii}(F_{pp})_{jj}},
& i\neq j.
\end{cases}
\end{equation}

Here, the exponential factor is the weight between bands $i$ and $j$ and decreases  with their separation $|k_i-k_j|$, while $k_{\rm corr}$ is the correlation scale and $\rho$ is the smoothing strength. The factor $\sqrt{(F_{pp})_{ii}(F_{pp})_{jj}}$  is a weight given by  the geometric mean of the information carried by bands $i$ and $j$. Clearly, we need to choose $\rho\ll 1$ to avoid saturating the Cauchy-Schwarz inequality and maintain positive-definiteness.

We will always quote fully marginalized errors at 1$\sigma$, as customary within the Fisher matrix approach, and discuss how a non-zero $\beta$ can be "distinguished" from the uncoupled case to 1$\sigma$. It is clear, however, that a robust detection can only be claimed if the confidence level is  better than, say, 3$\sigma$, so our uncertainties should be multiplied by three to reach this level.

\begin{figure}
    \centering
    \includegraphics[width=0.8\linewidth]{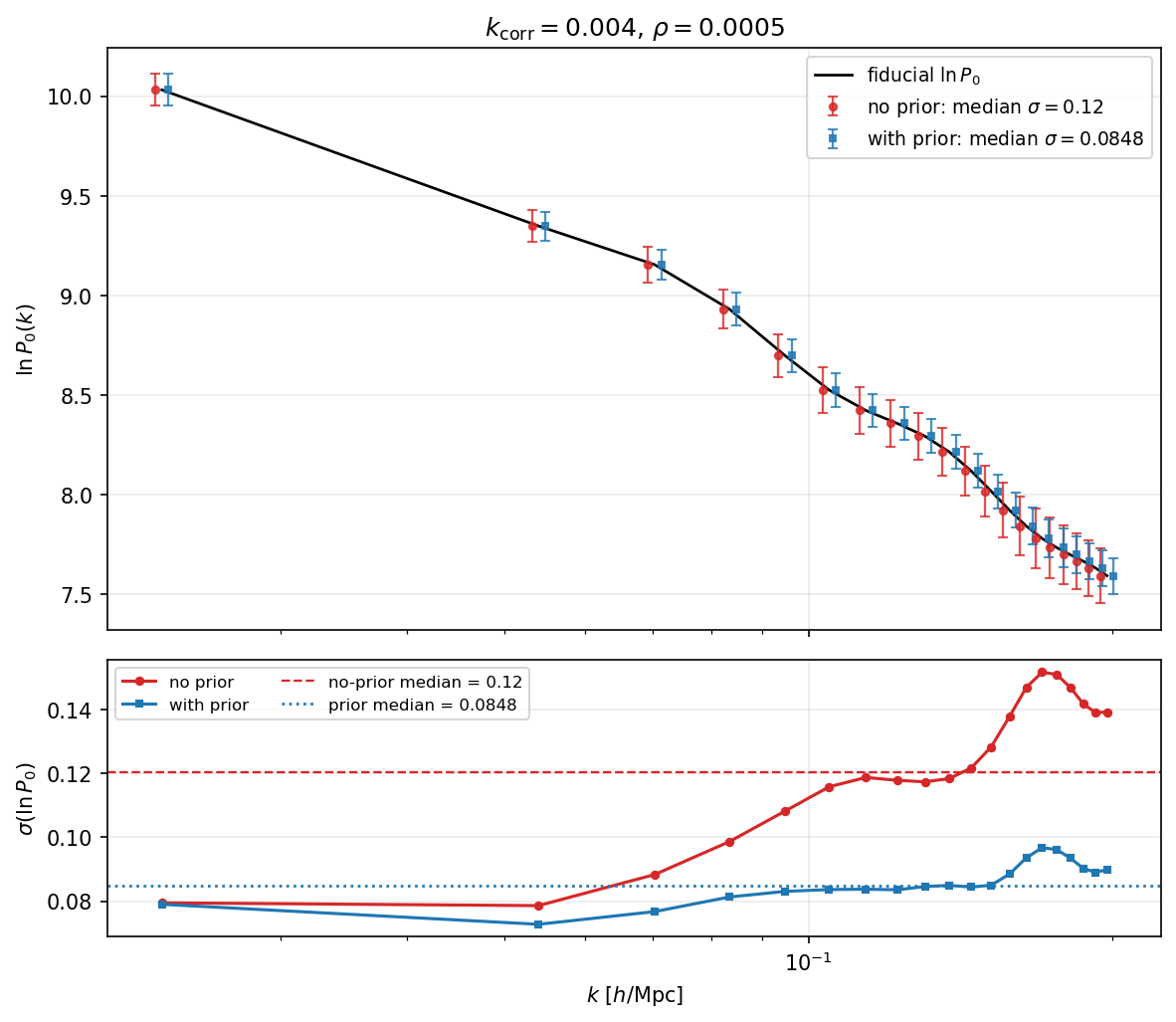}
    \caption{Effect of the Gaussian off-diagonal prior $\Pi_\rho$ on the power-spectrum bands for the joint DESI+Euclid forecast with $\beta_{\rm fid}=0.15$, $\nu=0.1$, and $k_{\max}=0.2\,h\,{\rm Mpc}^{-1}$, using $k_{\rm corr}=0.004\,h\,{\rm Mpc}^{-1}$ and $\rho=0.0005$. Top: fiducial $\ln P_0(k)$ and marginalized $1\sigma$ errors without and with the prior. Bottom: the corresponding band uncertainties, whose median decreases from $0.120$ without the prior to $0.0848$ with the prior. The two sets of error bars are slightly shifted horizontally for clarity; they correspond to the same $k$-band centers.}
    \label{fig:lnp0-errors}
\end{figure}

\begin{figure}
    \centering
    \includegraphics[width=1\linewidth]{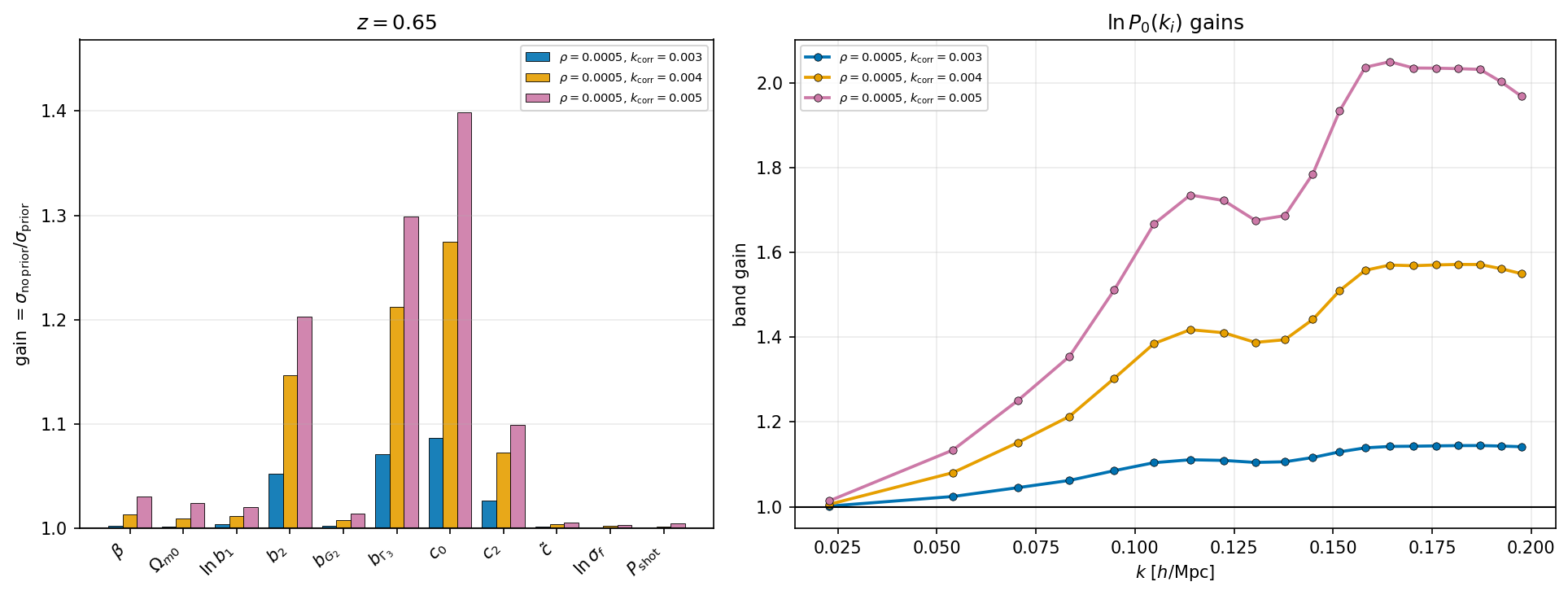}
    \caption{Effect of the Gaussian off-diagonal prior $\Pi_\rho$ on the joint DESI+Euclid forecast for $\beta_{\rm fid}=0.15$, $\nu=0.1$, and $k_{\max}=0.2\,h\,{\rm Mpc}^{-1}$. The gain is defined as $\sigma_{\rm no\,prior}/\sigma_{\rm prior}$. Left: gains for $\beta$, $\Omega_{m0}$, and the nuisance parameters in the lowest redshift bin, $z=0.65$. Right: gains for the individual $\ln P_0(k_i)$ bands. All errors are marginalized over the other parameters. We fix $\rho=0.0005$ and compare $k_{\rm corr}=0.003$, $0.004$, and $0.005\,h\,{\rm Mpc}^{-1}$. Values above unity indicate tighter constraints.}
    \label{fig:prior-gains}
\end{figure}

\begin{figure}
    \centering
    \includegraphics[width=1\linewidth]{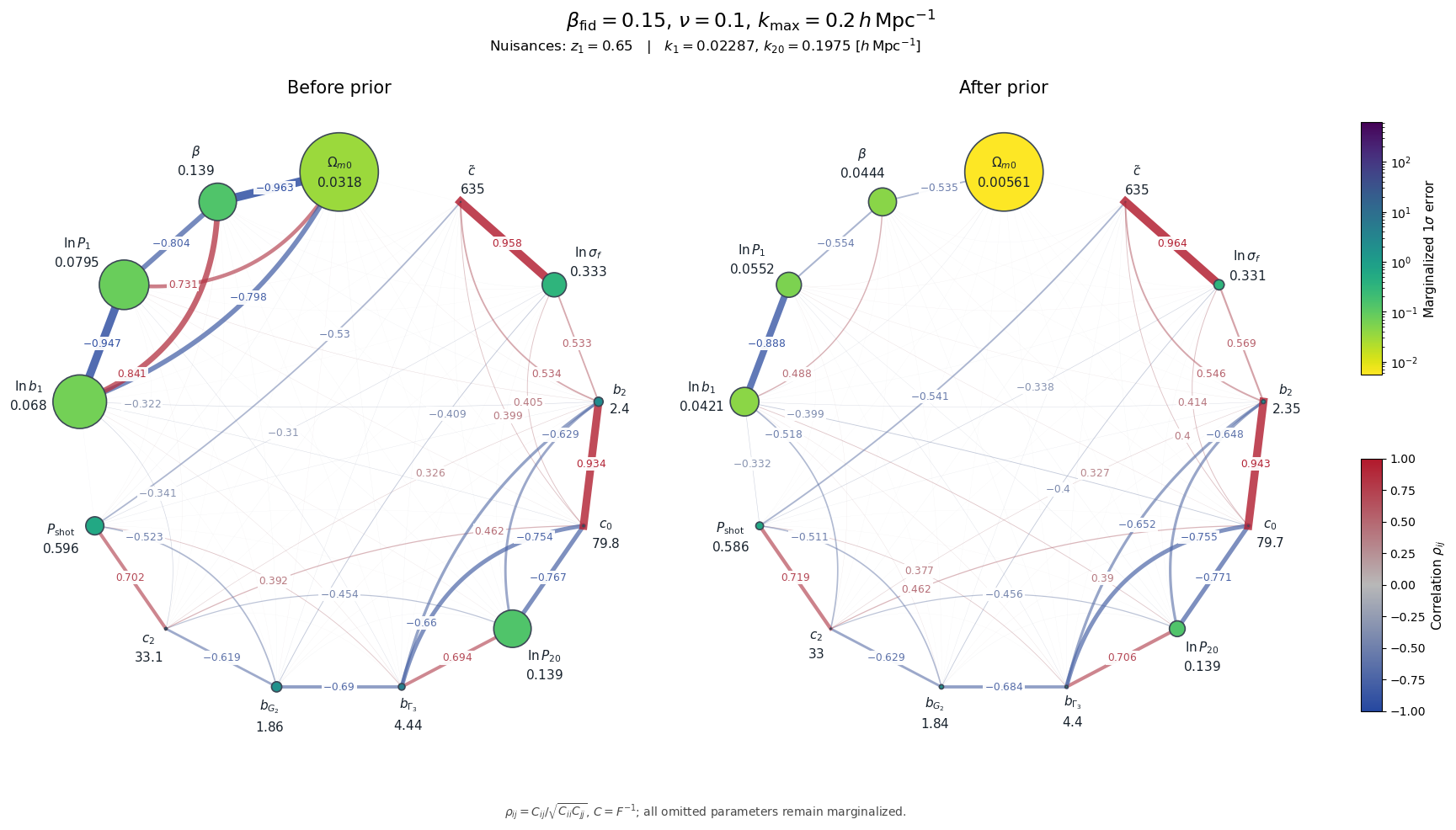}
    \caption{Parameter-correlation network for the joint DESI+Euclid forecast with $\beta_{\rm fid}=0.15$, $\nu=0.1$, and $k_{\max}=0.2\,h\,\mathrm{Mpc}^{-1}$. The nodes show the global parameters, the first and last $\ln P_0$ bands, and the nuisance parameters in the lowest-redshift bin ($z_1=0.65$); all omitted parameters remain marginalized. Node color encodes the marginalized $1\sigma$ error on a shared logarithmic scale, and node area scales as $1/\sigma$. Every pair of nodes is connected by a line: line colors show the signed marginalized correlation $\rho_{ij}=C_{ij}/\sqrt{C_{ii}C_{jj}}$, line widths scale with $|\rho_{ij}|$, and numerical labels are shown only for $|\rho_{ij}|\ge 0.3$. The left panel uses the original covariance, while the right panel includes the external Gaussian prior on $\Omega_{m0}$ with $\sigma_{\Omega,\mathrm{ext}}=0.0057$.}
    \label{fig:network}
\end{figure}

We adopt $k_{\rm corr}=0.004\,h\,{\rm Mpc}^{-1}$ and $\rho=0.0005$ as our reference values for most figures.  First, we verified that   each matrix element of the prior $|\Pi_{ij}|$ is small compared with $F_{pp}$ ( the maximum value of the ratio  $r_{ij} = \frac{|\Pi_{ij}|}{|(F_{pp})_{ij}|}$  is  $0.005$  ).  For our reference values, the median marginalized band uncertainty for the power spectrum decreases from $0.120$ to $0.0848$, giving a gain of approximately $1.42$ $(\frac{\mathrm{median}(\sigma)}{\mathrm{median}(\sigma_{\rm prior})})$, as illustrated in Figure~\ref{fig:lnp0-errors}. At fixed $\rho$, increasing $k_{\rm corr}$ from $0.003$ to $0.005\,h\,{\rm Mpc}^{-1}$ gives larger band-power gains  at non-linear scales,  with $c_0$ showing the strongest improvement among the nuisance parameters (Figure~\ref{fig:prior-gains}). We investigated why such a small prior can have a strong impact on the constraints. Two effects help explain this behaviour. Firstly, the Gaussian weight  decreases with band separation. Our binning places bands closer together at high $k$, giving neighbouring bands in the nonlinear regime larger prior weights. Secondly, there are several nuisance parameters  strongly correlated with band power at non-linear scales. These  degeneracies with nuisance parameters can  reduce the information remaining in the band powers, and amplify the response of the marginalized band errors. Therefore even a prior with small entries compared with $F_{pp}$ can become significant relative to this residual information (we discuss this effect  in Appendix~~\ref{app:prior-gains}).  For example, $c_{0}$ is the  parameter most strongly correlated with several nonlinear bands and has the largest gain (see left panel of Figure \ref{fig:prior-gains}). We verify that fixing $c_0$ substantially reduces the gain in band powers, and it is consistent with our interpretation (see Appendix~~\ref{app:prior-gains} and Figure \ref{fig:band_gain_fixed}).  For $\beta$, however, the improvement from the band prior is small because its correlations with the band powers on nonlinear scales are weak (see Figure~\ref{fig:network} for its correlation with the highest-$k$ band).

\begin{figure}
    \centering
    \includegraphics[width=1\linewidth]{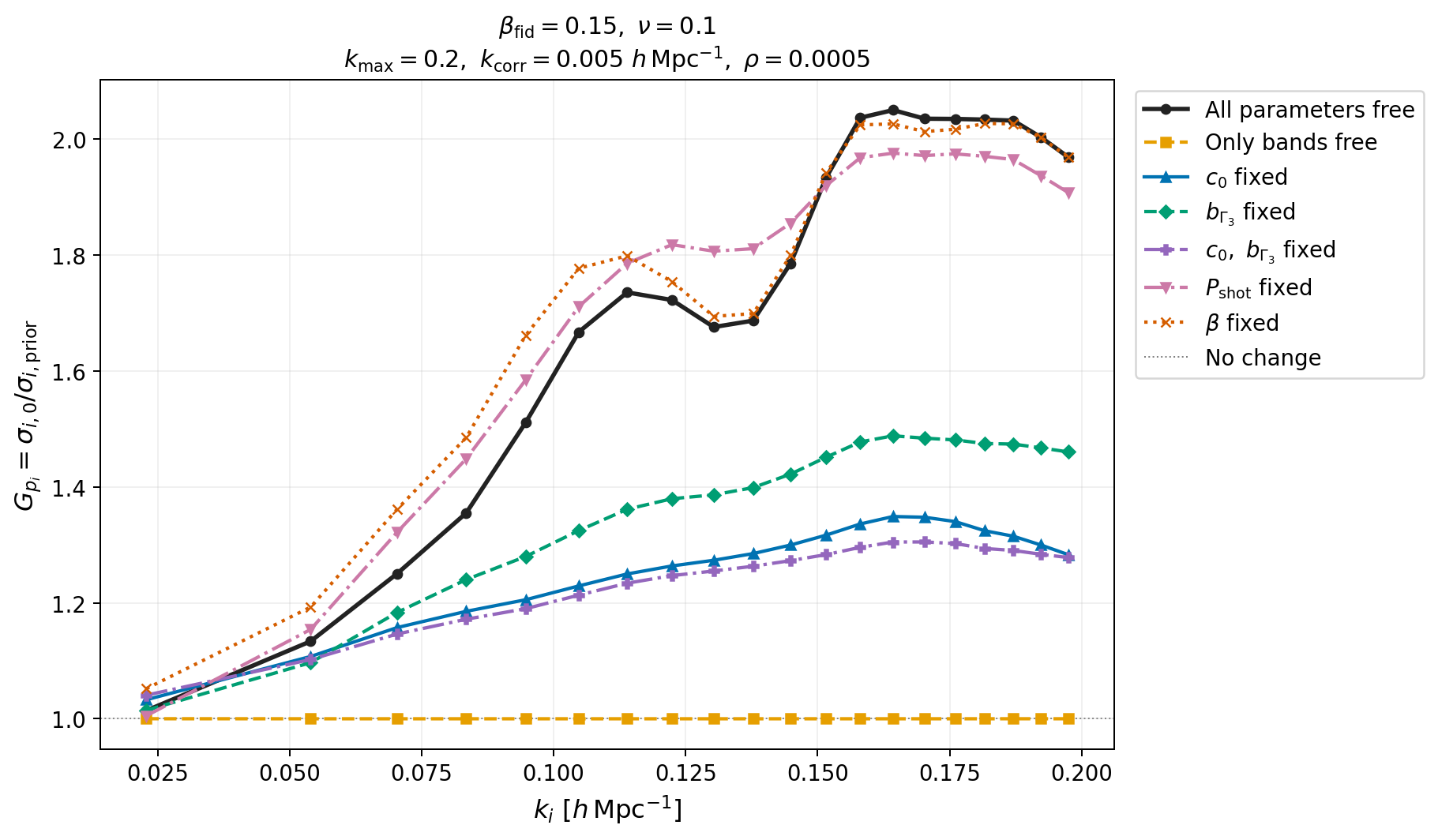}
    \caption{Gain in the marginalized $\ln P_0(k_i)$ band constraints for different fixed-parameter choices, using the Gaussian off-diagonal prior with $\rho=0.0005$, $k_{\rm corr}=0.005\,h\,{\rm Mpc}^{-1}$, and $k_{\max}=0.2\,h\,{\rm Mpc}^{-1}$. This diagnostic uses the joint DESI+Euclid forecast with $\beta_{\rm fid}=0.15$ and $\nu=0.1$. Curves compare all parameters free; only the band powers free; and $c_0$, $b_{\Gamma_3}$, both $c_0$ and $b_{\Gamma_3}$, $P_{\rm shot}$, or $\beta$ fixed. Each selected nuisance parameter is fixed in every redshift bin. The dotted line marks no change, $G_{p_i}=1$.}
    \label{fig:band_gain_fixed}
\end{figure}

\begin{figure}
    \centering
    \includegraphics[width=0.8\linewidth]{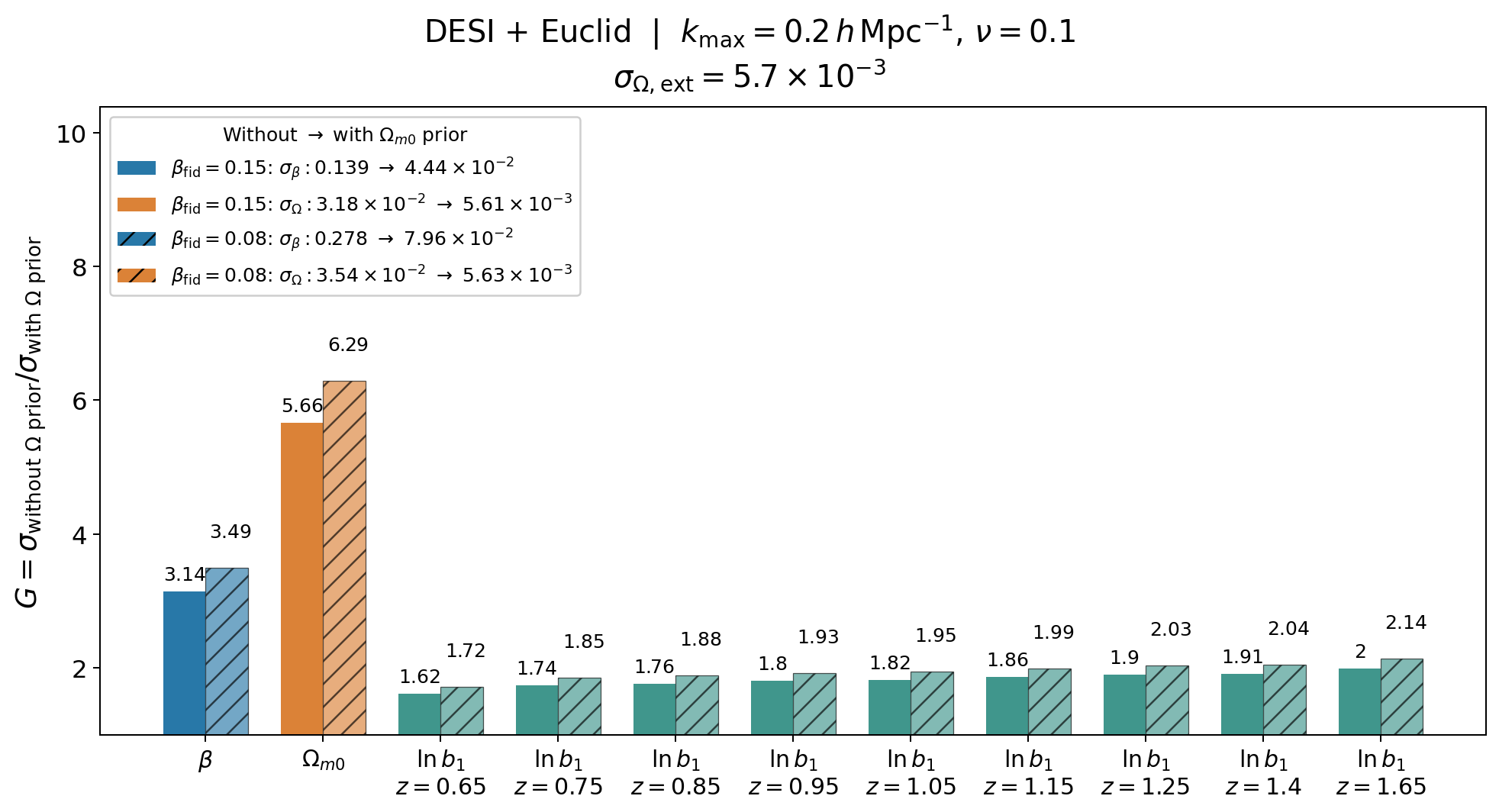}
    \caption{Effect of a Gaussian prior on $\Omega_{m0}$ with $\sigma_{\Omega,\mathrm{ext}}=0.0057$ on the DESI+Euclid forecast, for $\nu=0.1$ and $k_{\max}=0.2\,h\,\mathrm{Mpc}^{-1}$. Solid and hatched bars show $\beta_{\rm fid}=0.15$ and $0.08$, respectively. The gain is defined as $G=\sigma/\sigma_{\rm prior}$, with both errors marginalized over all other parameters. The bars show gains for $\beta$, $\Omega_{m0}$, and $\ln b_1$ in each redshift bin; the legend also reports the global constraints before and after applying the $\Omega_{m0}$ prior. No band-power prior is included.}
    \label{fig:omega_prior1}
\end{figure}

\begin{figure}
    \centering
    \includegraphics[width=0.8\linewidth]{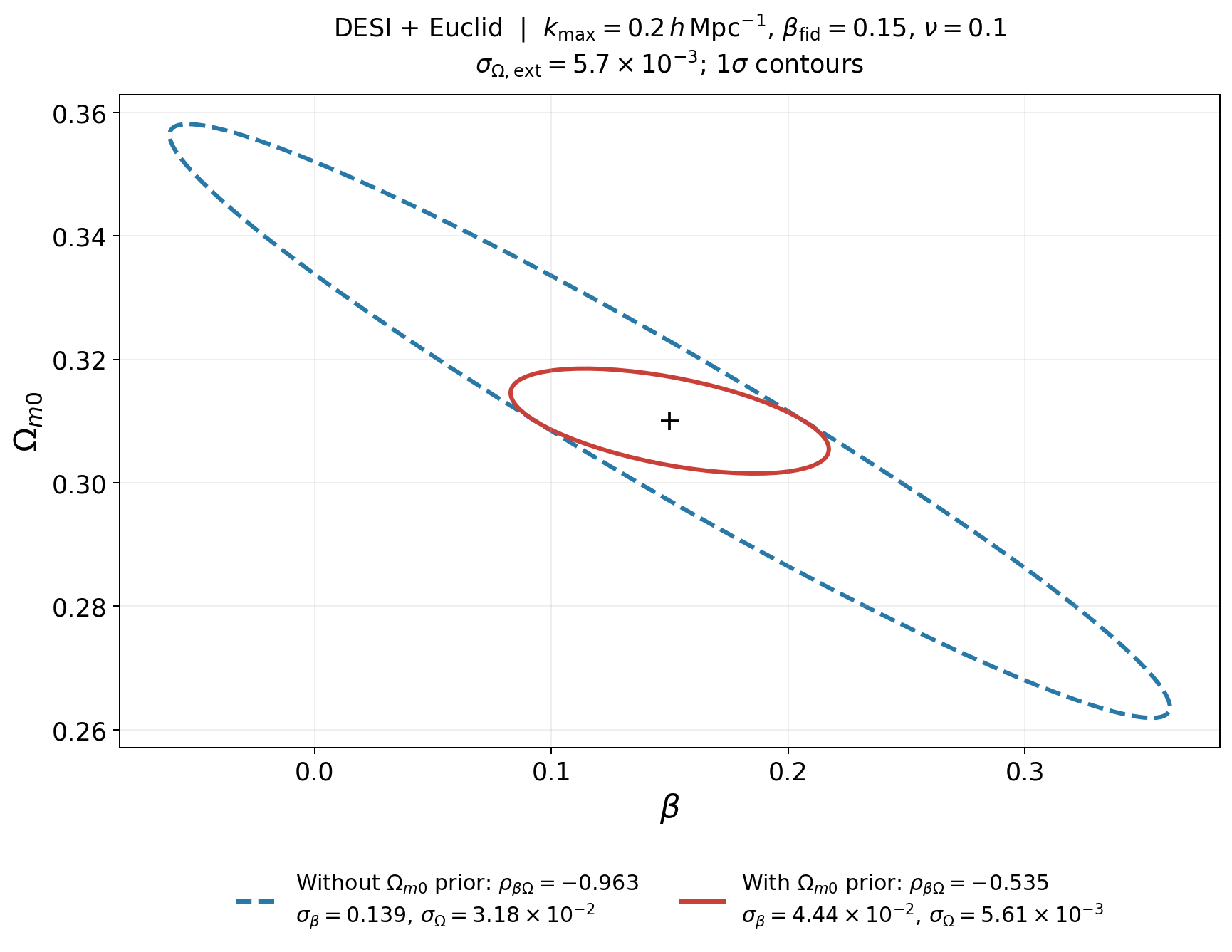}
   \caption{Marginalized $1\sigma$ contours in the $(\beta,\Omega_{m0})$ plane for DESI+Euclid, with $\beta_{\rm fid}=0.15$, $\nu=0.1$, and $k_{\max}=0.2\,h\,\mathrm{Mpc}^{-1}$. Blue dashed and red solid contours show the forecasts without and with a Gaussian prior on $\Omega_{m0}$, respectively. The prior is centred on $\Omega_{m0,\rm fid}=0.31$ with width $\sigma_{\Omega,\mathrm{ext}}=0.0057$. The legend reports the correlation coefficient $\rho_{\beta\Omega}$ and the marginalized one-parameter $1\sigma$ errors. No band-power prior is included.}
    \label{fig:omega_prior_elipses}
\end{figure}

As a further test, we investigate the degeneracy between $\beta$ and $\Omega_{m0}$. We apply a Gaussian prior centered on $\Omega_{m0,\rm fid}=0.31$ with width $\sigma_{\Omega,\rm ext}=0.0057$. This choice is motivated by the similar uncertainty of $0.0056$ obtained from Planck temperature, polarization, and lensing data combined with BAO measurements  (see Table 2 in \cite{Planck2020}), and by the uncertainty of $0.0057$ reported in Ref.~\cite{omori2026} from combined SPT-3G, Planck, and ACT primary-CMB and CMB-lensing data. We use the quoted uncertainty as an illustrative benchmark. For the joint DESI+Euclid forecast with $\beta_{\rm fid}=0.15$, $\nu=0.1$, and $k_{\max}=0.2\,h\,{\rm Mpc}^{-1}$,  the marginalized errors decrease from $0.139$ to $0.0444$ for $\beta$ and from $0.0318$ to $0.00561$ for $\Omega_{m0}$, giving gains of $3.14$ and $5.66$, respectively $\left(\frac{\sigma}{\sigma_{\rm prior}}\right)$ (Figure ~\ref{fig:omega_prior1}). Their anticorrelation weakens from $-0.963$ to $-0.535$  (as shown in Figures \ref{fig:network},   \ref{fig:omega_prior_elipses} ). Apart from the directly constrained $\Omega_{m0}$, the largest gain is in $\beta$, while $\ln b_1$ benefits most among the nuisance parameters, with gains of $1.62$--$2.00$ across redshift bins.  The gain in the other nuisance parameters is less than $3\%$. These improvements reflect the strong anticorrelation of $\beta$ with $\Omega_{m0}$ and the correlations of $\ln b_1$ with both parameters. Figure~\ref{fig:omega_prior1} also shows that, with $\beta_{\rm fid}=0.08$, the errors decrease from $0.278$ to $0.0796$ for $\beta$ and from $0.0354$ to $0.00563$ for $\Omega_{m0}$, giving gains of $3.49$ and $6.29$, respectively; the $\ln b_1$ gains range from $1.72$ to $2.14$.  With this prior information, $\beta_{\rm fid}/\sigma_\beta \simeq 3.38$ for $\beta_{\rm fid}=0.15$ and $\beta_{\rm fid}/\sigma_\beta \simeq 1.01$ for $\beta_{\rm fid}=0.08$. In the most optimistic case  with this prior and $k_{\rm max}=0.3 h/$Mpc, the smallest coupling distinguishable from zero at the $1\sigma$ level is $\beta_{\rm fid}=0.05$.

It is clear that by breaking the degeneracy with $\Omega_{m0}$ one can  substantially improve the constraints on the coupling. On the other hand, the CMB-motivated prior we employed has been obtained by assuming $\Lambda$CDM, which is of course inconsistent with our model. This  should be therefore taken just as an illustration of how much one can in principle push the bounds on $\beta$.

\subsection{Forecast Constraints}
We present  a joint DESI+Euclid forecast constructed from non-overlapping redshift bins (Table~\ref{tab:joint2-forecast-errors-nu-0p1})  and compare it with DESI and Euclid separately.  The tables in this section report forecasts without the band prior or the external Gaussian prior on $\Omega_{m0}$. Whenever either prior is included, it is explicitly stated in the caption. 

To improve constraints, we experiment with alternative combinations  and find that the best improvement is obtained when retaining all seven DESI bins at $z=0.65$, $0.75$, $0.85$, $0.95$, $1.05$, $1.15$, and $1.25$, and using the Euclid bins at $z=1.4$ and $1.65$. For $\beta_{\rm fid}=0.15$ and $\nu=0.1$, moving from the linear-regime scale cut $k_{\max}=0.10\,h\,{\rm Mpc}^{-1}$ to our reference value $k_{\max}=0.20\,h\,{\rm Mpc}^{-1}$ reduces $\sigma(\beta)$ from $0.692$ to $0.139$ and $\sigma(\Omega_{m0})$ from $0.175$ to $0.0318$, giving gains of $4.96$ and $5.51$, respectively $\left(\frac{\sigma_{0.10}}{\sigma_{0.20}}\right)$. At the reference scale cut, the joint forecast improves the constraints relative to DESI alone by factors of $1.60$ for $\beta$ and $1.39$ for $\Omega_{m0}$ $\left(\frac{\sigma_{\rm DESI}}{\sigma_{\rm Joint}}\right)$. Relative to Euclid alone, the corresponding gains are $1.58$ and $1.67$ $\left(\frac{\sigma_{\rm Euclid}}{\sigma_{\rm Joint}}\right)$.

\begin{figure}
    \centering
    \includegraphics[width=1.\linewidth]{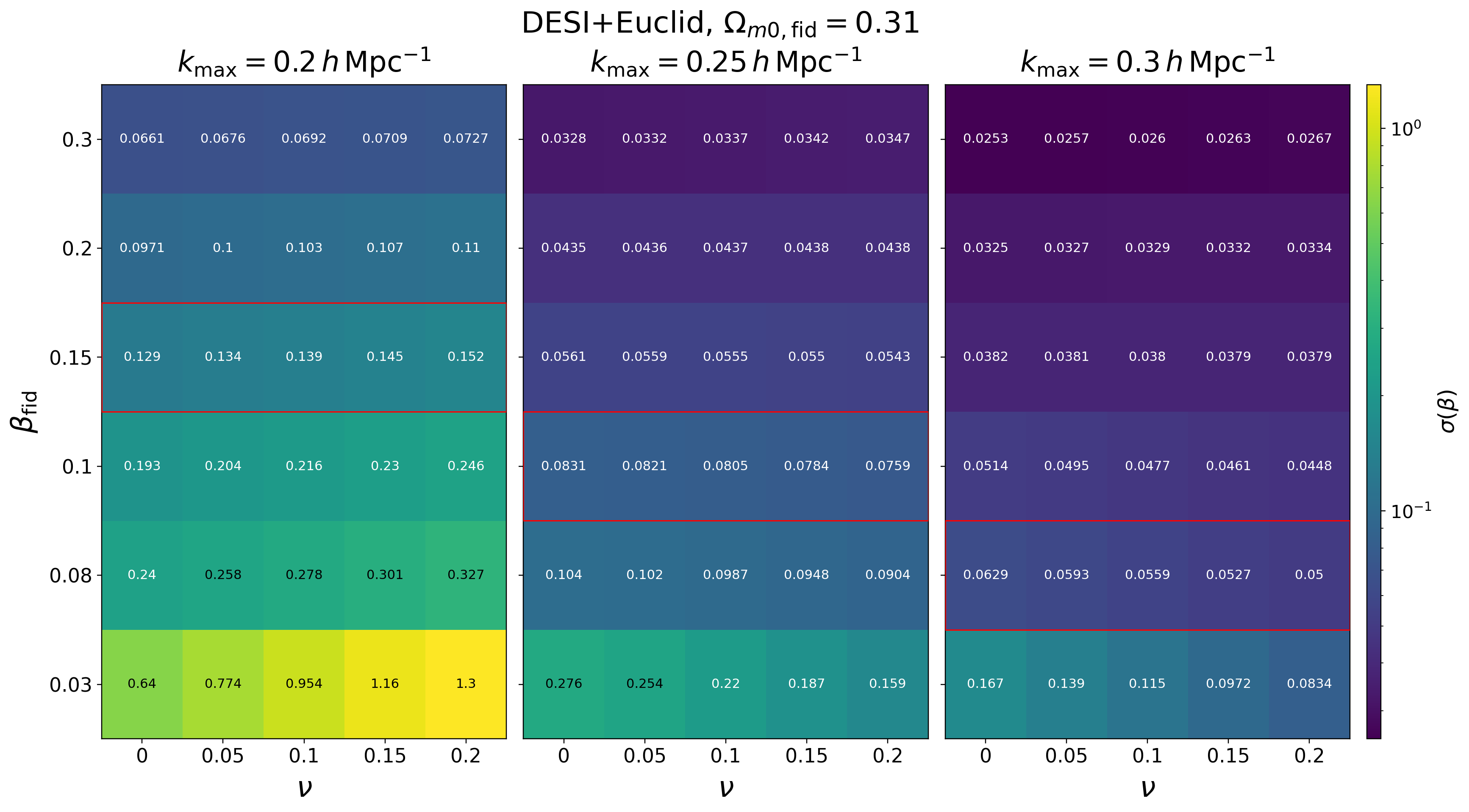}
    \caption{Marginalized constraint on the coupling $\beta$ for the joint DESI+Euclid forecast as a function of the fiducial coupling $\beta_{ fid}$ and the fixed potential-slope parameter $\nu$, with $\Omega_{m0,\rm fid}=0.31$. From left to right, the panels show $k_{\max}=0.20$, $0.25$, and $0.30\,h\,{\rm Mpc}^{-1}$; the numerical values and color scale give $\sigma(\beta)$. The red rectangles highlight the lowest sampled $\beta_{\rm fid}$ satisfying $\sigma(\beta)<\beta_{\rm fid}$ at $\nu=0.1$ for each scale cut: $0.15$, $0.10$, and $0.08$, respectively. These values mark the sampled $1\sigma$ distinguishability thresholds at this slope.}
\label{fig:beta_nu_constraints}
\end{figure}

\begin{figure}
    \centering
    \includegraphics[width=1\linewidth]{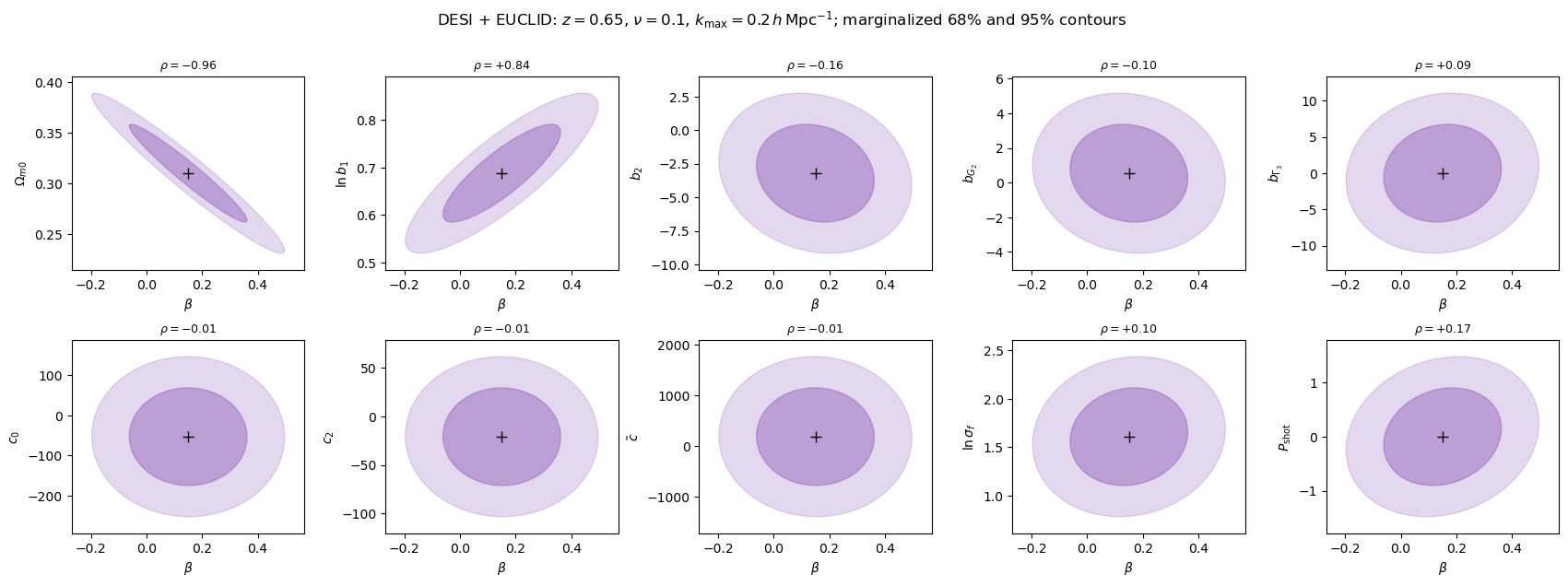}
    \caption{Marginalized $68\%$ and $95\%$ confidence contours between $\beta$ and $\Omega_{m0}$ or each nuisance parameter in the lowest joint survey redshift bin, $z=0.65$, for fixed $\nu=0.1$, $k_{\max}=0.2\,h\,{\rm Mpc}^{-1}$ and $\beta_{fid}=0.15$. Dark and light regions denote the $68\%$ and $95\%$ contours, respectively, while crosses mark the fiducial values. The correlation coefficient $\rho$ is reported above each panel.}
    \label{fig:desi-beta-correlations}
\end{figure}

\begin{figure}
    \centering
    \includegraphics[width=0.8\linewidth]{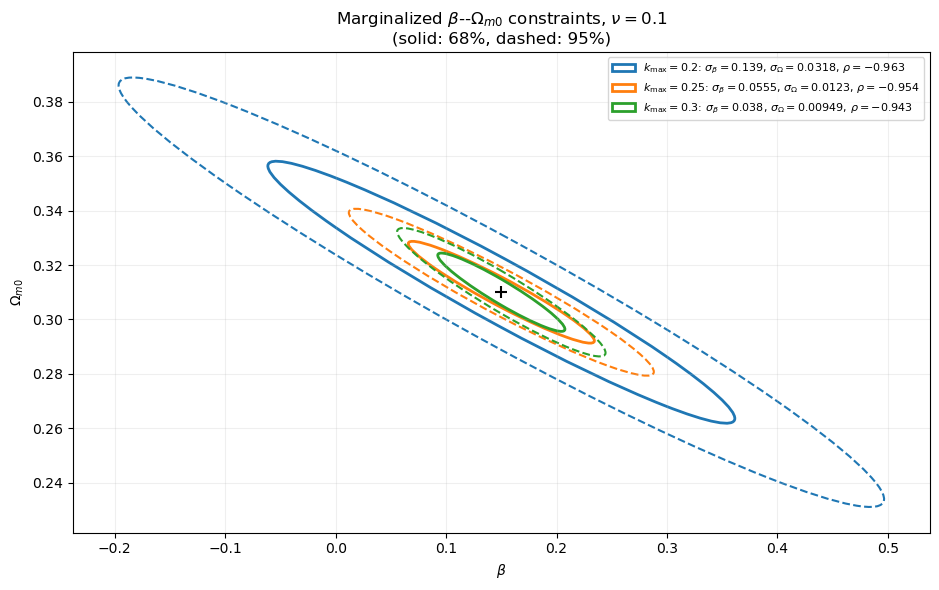}
    \caption{ Marginalized $68\%$ and $95\%$ confidence contours in the $\beta$--$\Omega_{m0}$ plane for the  joint  (DESI + Euclid) forecast with fixed $\nu=0.1$ and $\beta_{fid}=0.15$. Colors denote the scale cuts $k_{\max}=0.20$, $0.25$, and $0.30\,h\,{\rm Mpc}^{-1}$; solid and dashed curves show the $68\%$ and $95\%$ contours, respectively, and the cross marks the fiducial point. The legend reports the marginalized uncertainties and correlation coefficient for each case.}
    \label{fig:beta-omega-kmax-contours}
\end{figure}

\begin{figure}
    \centering
    \includegraphics[width=0.8\linewidth]{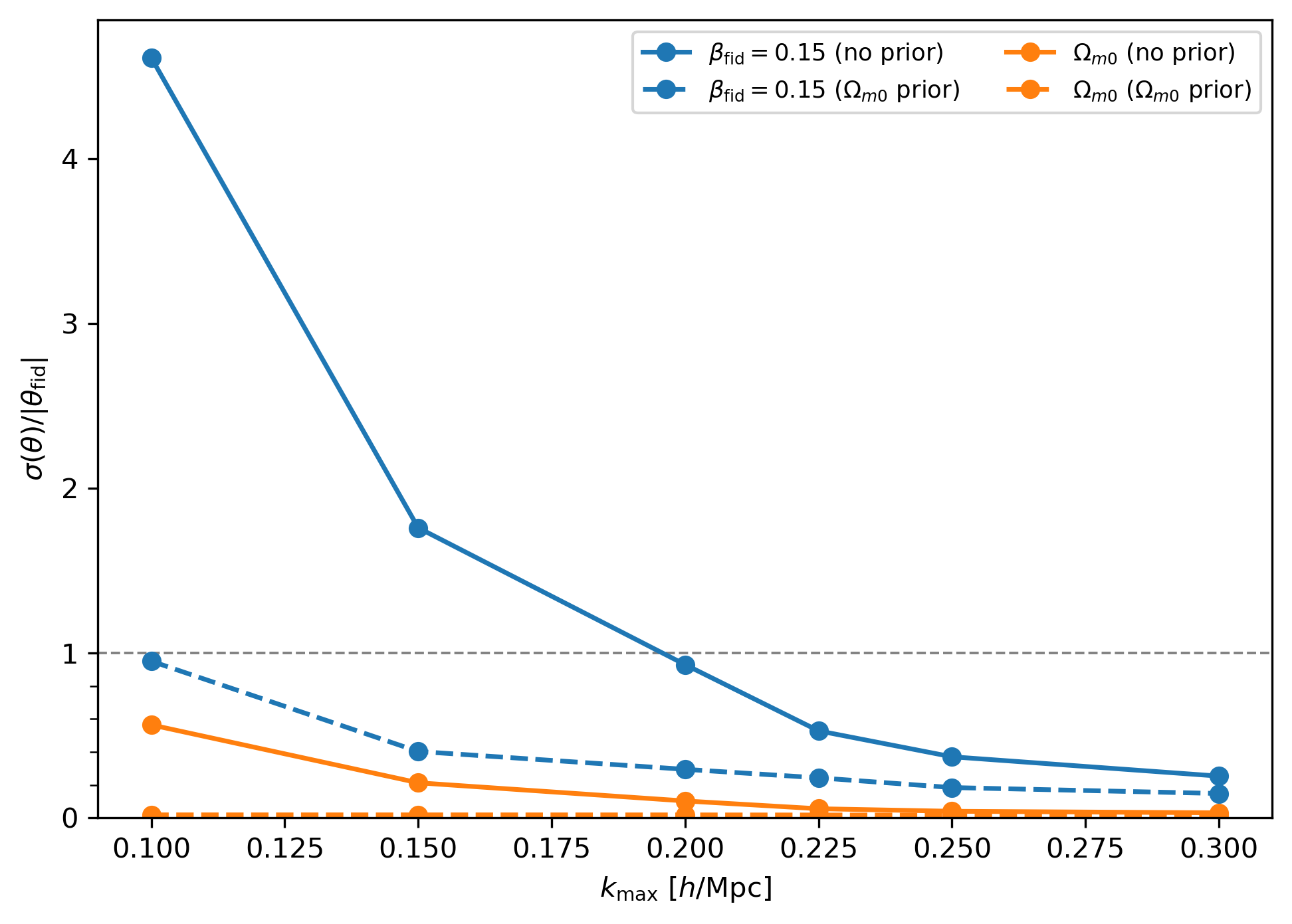}
  \caption{Marginalized $1\sigma$ relative errors on $\beta$ (blue) and $\Omega_{m0}$ (orange) as functions of the scale cut $k_{\max}$ for the joint DESI+Euclid forecast, with fixed $\nu=0.1$ and $\beta_{\rm fid}=0.15$. Solid curves show forecasts without an external $\Omega_{m0}$ prior, while dashed curves show forecasts with the Gaussian prior $\sigma_{\Omega,\rm ext}=0.0057$. The horizontal gray line marks $\sigma(\beta)/\beta_{\rm fid}=1$; below this line, the  coupling can be distinguished from $\beta=0$ at the $1\sigma$ level or higher. 
   }
    \label{fig:kmax-constraints}
\end{figure}

We consider now how the  constraint on $\beta$ depends on the fiducial coupling, the potential slope, and the scale cut. We find that, at $\nu=0.1$, the lowest sampled coupling parameters distinguishable from zero at the $1\sigma$ level are at $\beta_{\rm fid}=0.15$, $0.10$, and $0.08$ for $k_{\max}=0.20$, $0.25$, and $0.30\,h\,{\rm Mpc}^{-1}$, respectively (Figure~\ref{fig:beta_nu_constraints}). For $\beta_{\rm fid}=0.15$ and $k_{\max}=0.20\,h\,{\rm Mpc}^{-1}$, increasing $\nu$ from $0$ to $0.2$ raises $\sigma(\beta)$ from $0.129$ to $0.152$, a degradation factor of $1.18$ $\left(\frac{\sigma_{\nu=0.2}}{\sigma_{\nu=0}}\right)$. At the reference slope $\nu=0.1$, DESI and Euclid give similar coupling errors, $\sigma(\beta)=0.223$ and $0.220$, while their errors on $\Omega_{m0}$ are $0.0441$ and $0.0532$, respectively (Tables~\ref{tab:desi-forecast-errors-nu-0p1} and~\ref{tab:euclid-forecast-errors-nu-0p1}). Neither survey alone therefore distinguishes $\beta_{\rm fid}=0.15$ from zero at $1\sigma$ for this scale cut. The joint forecast retains a strong $\beta$--$\Omega_{m0}$ anticorrelation, $\rho\simeq-0.963$ (Figure~\ref{fig:desi-beta-correlations}). Including smaller scales substantially tightens the constraints: increasing $k_{\max}$ from $0.20$ to $0.30\,h\,{\rm Mpc}^{-1}$ reduces the joint errors from $0.139$ to $0.0380$ for $\beta$ and from $0.0318$ to $0.00949$ for $\Omega_{m0}$, giving gains of $3.67$ and $3.34$, respectively $\left(\frac{\sigma_{0.20}}{\sigma_{0.30}}\right)$ (Figures~\ref{fig:beta-omega-kmax-contours} and~\ref{fig:kmax-constraints}).

\begin{table*}[t]
  \centering
  \caption{Marginalized $1\sigma$ forecast errors for the joint DESI+Euclid forecast at $\nu=0.1$ and $\beta_{\rm fid}=0.15$. Each scale-cut section reports the global constraints, the $\ln P_0$ error range, and nuisance constraints in each redshift bin.}
  \label{tab:joint2-forecast-errors-nu-0p1}
  {\scriptsize
  \setlength{\tabcolsep}{3.5pt}
  \renewcommand{\arraystretch}{1.05}
  \begin{tabular}{crrrrrrrrr}
    \hline
    \multicolumn{10}{|c|}{\textbf{Joint (DESI + EUCLID) Forecast}} \\
    \hline
    \multicolumn{10}{|c|}{\textbf{$k_{\max}=0.1\,h\,\mathrm{Mpc}^{-1}$}} \\
    \multicolumn{10}{|c|}{$\sigma(\beta)=6.92\times 10^{-1}$ ($\frac{\sigma(\beta)}{\beta_{\rm fid}}=4.611$)\hspace{1em}$\sigma(\Omega_{m0})=1.75\times 10^{-1}$ ($\frac{\sigma(\Omega_{m0})}{\Omega_{m0,\rm fid}}=0.564$)} \\
    \multicolumn{10}{|c|}{$\sigma(\ln P_0)$: min 0.269, median 0.783, max 1.68} \\
    \hline
    $z$ & $\ln b_1$ & $b_2$ & $b_{G_2}$ & $b_{\Gamma_3}$ & $c_0$ & $c_2$ & $\tilde c$ & $\ln\sigma_f$ & $P_{\rm shot}$ \\
    \hline
    0.65 & 0.284 & 27 & 14 & 24.3 & 849 & 526 & $7.15\times 10^{3}$ & 0.77 & 0.986 \\
    0.75 & 0.302 & 28.5 & 13.8 & 26.2 & 807 & 489 & $5.86\times 10^{3}$ & 0.729 & 0.996 \\
    0.85 & 0.32 & 31.6 & 14.3 & 28.8 & 788 & 476 & $5.18\times 10^{3}$ & 0.705 & 0.99 \\
    0.95 & 0.337 & 35.5 & 15 & 31.6 & 774 & 466 & $4.69\times 10^{3}$ & 0.688 & 0.98 \\
    1.05 & 0.355 & 40.5 & 16.3 & 35.2 & 773 & 470 & $4.49\times 10^{3}$ & 0.697 & 0.934 \\
    1.15 & 0.372 & 45.8 & 17.5 & 38.8 & 771 & 469 & $4.29\times 10^{3}$ & 0.705 & 0.897 \\
    1.25 & 0.389 & 51.6 & 18.8 & 42.7 & 770 & 469 & $4.13\times 10^{3}$ & 0.715 & 0.865 \\
    1.4 & 0.42 & 50.3 & 12.4 & 37.7 & 535 & 311 & $2.93\times 10^{3}$ & 0.708 & 0.685 \\
    1.65 & 0.467 & 66.2 & 14.6 & 46.8 & 540 & 314 & $2.8\times 10^{3}$ & 0.725 & 0.44 \\
    \hline
    \multicolumn{10}{|c|}{\textbf{$k_{\max}=0.2\,h\,\mathrm{Mpc}^{-1}$}} \\
    \multicolumn{10}{|c|}{$\sigma(\beta)=1.39\times 10^{-1}$ ($\frac{\sigma(\beta)}{\beta_{\rm fid}}=0.929$)\hspace{1em}$\sigma(\Omega_{m0})=3.18\times 10^{-2}$ ($\frac{\sigma(\Omega_{m0})}{\Omega_{m0,\rm fid}}=0.102$)} \\
    \multicolumn{10}{|c|}{$\sigma(\ln P_0)$: min 0.0786, median 0.12, max 0.152} \\
    \hline
    $z$ & $\ln b_1$ & $b_2$ & $b_{G_2}$ & $b_{\Gamma_3}$ & $c_0$ & $c_2$ & $\tilde c$ & $\ln\sigma_f$ & $P_{\rm shot}$ \\
    \hline
    0.65 & 0.068 & 2.4 & 1.86 & 4.44 & 79.8 & 33.1 & 635 & 0.333 & 0.596 \\
    0.75 & 0.0709 & 2.61 & 1.7 & 4.66 & 79.2 & 26.2 & 492 & 0.289 & 0.798 \\
    0.85 & 0.0739 & 2.94 & 1.92 & 5.21 & 80.4 & 29.6 & 480 & 0.29 & 0.627 \\
    0.95 & 0.077 & 3.29 & 2.11 & 5.73 & 82 & 31.9 & 456 & 0.286 & 0.519 \\
    1.05 & 0.0804 & 3.72 & 2.47 & 6.37 & 84.5 & 36.8 & 466 & 0.301 & 0.336 \\
    1.15 & 0.0836 & 4.16 & 2.73 & 6.99 & 86.4 & 39.1 & 454 & 0.303 & 0.286 \\
    1.25 & 0.0866 & 4.62 & 3 & 7.64 & 88.1 & 41.1 & 443 & 0.302 & 0.26 \\
    1.4 & 0.0922 & 4.62 & 2.05 & 7.2 & 74 & 28 & 297 & 0.206 & 0.203 \\
    1.65 & 0.1 & 5.81 & 2.58 & 8.8 & 76.5 & 30.7 & 296 & 0.212 & 0.123 \\
    \hline
    \multicolumn{10}{|c|}{\textbf{$k_{\max}=0.25\,h\,\mathrm{Mpc}^{-1}$}} \\
    \multicolumn{10}{|c|}{$\sigma(\beta)=5.55\times 10^{-2}$ ($\frac{\sigma(\beta)}{\beta_{\rm fid}}=0.37$)\hspace{1em}$\sigma(\Omega_{m0})=1.23\times 10^{-2}$ ($\frac{\sigma(\Omega_{m0})}{\Omega_{m0,\rm fid}}=0.04$)} \\
    \multicolumn{10}{|c|}{$\sigma(\ln P_0)$: min 0.0427, median 0.0625, max 0.0695} \\
    \hline
    $z$ & $\ln b_1$ & $b_2$ & $b_{G_2}$ & $b_{\Gamma_3}$ & $c_0$ & $c_2$ & $\tilde c$ & $\ln\sigma_f$ & $P_{\rm shot}$ \\
    \hline
    0.65 & 0.0338 & 0.762 & 1.11 & 2.16 & 28.6 & 16.8 & 263 & 0.137 & 0.225 \\
    0.75 & 0.034 & 0.849 & 1.06 & 2.31 & 28.8 & 13.9 & 205 & 0.116 & 0.403 \\
    0.85 & 0.0353 & 1 & 1.21 & 2.64 & 30.6 & 14.2 & 204 & 0.127 & 0.265 \\
    0.95 & 0.0363 & 1.17 & 1.35 & 2.98 & 32.3 & 14.4 & 202 & 0.136 & 0.21 \\
    1.05 & 0.0376 & 1.38 & 1.61 & 3.46 & 34.5 & 16.1 & 226 & 0.162 & 0.134 \\
    1.15 & 0.0387 & 1.58 & 1.83 & 3.93 & 36.1 & 17 & 233 & 0.174 & 0.119 \\
    1.25 & 0.0397 & 1.8 & 2.04 & 4.42 & 37.5 & 18 & 236 & 0.183 & 0.111 \\
    1.4 & 0.041 & 1.81 & 1.44 & 4.1 & 32.1 & 11.6 & 179 & 0.139 & 0.0803 \\
    1.65 & 0.0434 & 2.33 & 1.85 & 5.2 & 34 & 13.1 & 179 & 0.144 & 0.0502 \\
    \hline
    \multicolumn{10}{|c|}{\textbf{$k_{\max}=0.3\,h\,\mathrm{Mpc}^{-1}$}} \\
    \multicolumn{10}{|c|}{$\sigma(\beta)=3.8\times 10^{-2}$ ($\frac{\sigma(\beta)}{\beta_{\rm fid}}=0.253$)\hspace{1em}$\sigma(\Omega_{m0})=9.49\times 10^{-3}$ ($\frac{\sigma(\Omega_{m0})}{\Omega_{m0,\rm fid}}=0.031$)} \\
    \multicolumn{10}{|c|}{$\sigma(\ln P_0)$: min 0.0324, median 0.0368, max 0.0412} \\
    \hline
    $z$ & $\ln b_1$ & $b_2$ & $b_{G_2}$ & $b_{\Gamma_3}$ & $c_0$ & $c_2$ & $\tilde c$ & $\ln\sigma_f$ & $P_{\rm shot}$ \\
    \hline
    0.65 & 0.0263 & 0.421 & 0.585 & 1.3 & 16.8 & 7.87 & 134 & 0.07 & 0.0791 \\
    0.75 & 0.0266 & 0.458 & 0.568 & 1.43 & 16.7 & 6.43 & 97.7 & 0.0553 & 0.151 \\
    0.85 & 0.0275 & 0.555 & 0.643 & 1.66 & 18.2 & 6.59 & 103 & 0.0635 & 0.103 \\
    0.95 & 0.0284 & 0.658 & 0.725 & 1.91 & 19.6 & 6.9 & 106 & 0.0702 & 0.0861 \\
    1.05 & 0.0293 & 0.793 & 0.886 & 2.25 & 21.3 & 8.31 & 129 & 0.0909 & 0.0604 \\
    1.15 & 0.0301 & 0.921 & 1.02 & 2.57 & 22.6 & 9.24 & 137 & 0.101 & 0.0562 \\
    1.25 & 0.0308 & 1.05 & 1.16 & 2.92 & 23.6 & 10.1 & 141 & 0.108 & 0.0544 \\
    1.4 & 0.0319 & 1.05 & 0.923 & 2.83 & 20.2 & 8.06 & 114 & 0.0872 & 0.0398 \\
    1.65 & 0.0333 & 1.36 & 1.21 & 3.61 & 21.5 & 9.24 & 116 & 0.0925 & 0.0264 \\
    \hline
  \end{tabular}
  }
\end{table*}

\begin{table*}[t]
  \centering
  \caption{Marginalized $1\sigma$ forecast errors for the DESI survey at $\nu=0.1$ and $ \beta_{\rm fid}=0.15$. Each scale-cut section reports the global constraints, the $\ln P_0$ error range, and nuisance constraints.}
  \label{tab:desi-forecast-errors-nu-0p1}
  {\scriptsize
  \setlength{\tabcolsep}{3.5pt}
  \renewcommand{\arraystretch}{1.05}
  \begin{tabular}{crrrrrrrrr}
    \hline
    \multicolumn{10}{|c|}{\textbf{DESI survey}} \\
    \hline
    \multicolumn{10}{|c|}{\textbf{$k_{\max}=0.2\,h\,\mathrm{Mpc}^{-1}$}} \\
    \multicolumn{10}{|c|}{$\sigma(\beta)=2.23\times 10^{-1}$ ($\frac{\sigma(\beta)}{\beta_{\rm fid}}=1.486$)\hspace{1em}$\sigma(\Omega_{m0})=4.41\times 10^{-2}$ ($\frac{\sigma(\Omega_{m0})}{\Omega_{m0,\rm fid}}=0.142$)} \\
    \multicolumn{10}{|c|}{$\sigma(\ln P_0)$: min 0.142, median 0.197, max 0.236} \\
    \hline
    $z$ & $\ln b_1$ & $b_2$ & $b_{G_2}$ & $b_{\Gamma_3}$ & $c_0$ & $c_2$ & $\tilde c$ & $\ln\sigma_f$ & $P_{\rm shot}$ \\
    \hline
    0.65 & 0.111 & 3.29 & 2.2 & 7.03 & 125 & 37 & 652 & 0.337 & 0.625 \\
    0.75 & 0.117 & 3.73 & 2.11 & 7.72 & 128 & 31.1 & 509 & 0.293 & 0.816 \\
    0.85 & 0.121 & 4.27 & 2.36 & 8.61 & 132 & 35.5 & 498 & 0.295 & 0.641 \\
    0.95 & 0.126 & 4.86 & 2.57 & 9.51 & 136 & 39 & 476 & 0.291 & 0.531 \\
    1.05 & 0.131 & 5.54 & 2.89 & 10.5 & 140 & 44.8 & 488 & 0.307 & 0.346 \\
    1.15 & 0.136 & 6.26 & 3.14 & 11.5 & 144 & 48.2 & 480 & 0.31 & 0.299 \\
    1.25 & 0.141 & 7.02 & 3.4 & 12.5 & 147 & 51.1 & 472 & 0.312 & 0.276 \\
    \hline
    \multicolumn{10}{|c|}{\textbf{$k_{\max}=0.25\,h\,\mathrm{Mpc}^{-1}$}} \\
    \multicolumn{10}{|c|}{$\sigma(\beta)=6.77\times 10^{-2}$ ($\frac{\sigma(\beta)}{\beta_{\rm fid}}=0.451$)\hspace{1em}$\sigma(\Omega_{m0})=1.46\times 10^{-2}$ ($\frac{\sigma(\Omega_{m0})}{\Omega_{m0,\rm fid}}=0.047$)} \\
    \multicolumn{10}{|c|}{$\sigma(\ln P_0)$: min 0.0653, median 0.0786, max 0.0968} \\
    \hline
    $z$ & $\ln b_1$ & $b_2$ & $b_{G_2}$ & $b_{\Gamma_3}$ & $c_0$ & $c_2$ & $\tilde c$ & $\ln\sigma_f$ & $P_{\rm shot}$ \\
    \hline
    0.65 & 0.0495 & 0.973 & 1.28 & 3.03 & 41.1 & 18.9 & 286 & 0.141 & 0.257 \\
    0.75 & 0.0503 & 1.12 & 1.25 & 3.36 & 42.2 & 15.6 & 225 & 0.12 & 0.451 \\
    0.85 & 0.0519 & 1.33 & 1.42 & 3.85 & 44.5 & 15.6 & 218 & 0.13 & 0.283 \\
    0.95 & 0.0534 & 1.56 & 1.58 & 4.35 & 46.7 & 15.6 & 212 & 0.138 & 0.217 \\
    1.05 & 0.055 & 1.82 & 1.84 & 4.98 & 49.1 & 17 & 234 & 0.164 & 0.136 \\
    1.15 & 0.0563 & 2.09 & 2.07 & 5.61 & 51 & 18 & 240 & 0.177 & 0.12 \\
    1.25 & 0.0576 & 2.37 & 2.3 & 6.26 & 52.8 & 18.9 & 243 & 0.186 & 0.113 \\
    \hline
    \multicolumn{10}{|c|}{\textbf{$k_{\max}=0.3\,h\,\mathrm{Mpc}^{-1}$}} \\
    \multicolumn{10}{|c|}{$\sigma(\beta)=4.22\times 10^{-2}$ ($\frac{\sigma(\beta)}{\beta_{\rm fid}}=0.281$)\hspace{1em}$\sigma(\Omega_{m0})=1.05\times 10^{-2}$ ($\frac{\sigma(\Omega_{m0})}{\Omega_{m0,\rm fid}}=0.034$)} \\
    \multicolumn{10}{|c|}{$\sigma(\ln P_0)$: min 0.0385, median 0.0498, max 0.0621} \\
    \hline
    $z$ & $\ln b_1$ & $b_2$ & $b_{G_2}$ & $b_{\Gamma_3}$ & $c_0$ & $c_2$ & $\tilde c$ & $\ln\sigma_f$ & $P_{\rm shot}$ \\
    \hline
    0.65 & 0.0369 & 0.518 & 0.68 & 2 & 24.3 & 8.08 & 150 & 0.0722 & 0.0837 \\
    0.75 & 0.0376 & 0.585 & 0.708 & 2.29 & 25 & 6.66 & 113 & 0.0568 & 0.161 \\
    0.85 & 0.0388 & 0.71 & 0.808 & 2.65 & 26.9 & 6.79 & 114 & 0.0644 & 0.108 \\
    0.95 & 0.0399 & 0.844 & 0.917 & 3.04 & 28.5 & 7.09 & 114 & 0.0708 & 0.0899 \\
    1.05 & 0.0409 & 1 & 1.09 & 3.5 & 30.4 & 8.49 & 135 & 0.0916 & 0.0627 \\
    1.15 & 0.0418 & 1.16 & 1.25 & 3.96 & 31.8 & 9.45 & 142 & 0.102 & 0.0582 \\
    1.25 & 0.0427 & 1.33 & 1.4 & 4.45 & 33 & 10.4 & 146 & 0.109 & 0.0563 \\
    \hline
  \end{tabular}
  }
\end{table*}

\begin{table*}[t]
  \centering
  \caption{Marginalized $1\sigma$ forecast errors for the Euclid survey at $\nu=0.1$ and $ \beta_{\rm fid}=0.15$. Each scale-cut section reports the global constraints, the $\ln P_0$ error range, and nuisance constraints.}
  \label{tab:euclid-forecast-errors-nu-0p1}
  {\scriptsize
  \setlength{\tabcolsep}{3.5pt}
  \renewcommand{\arraystretch}{1.05}
  \begin{tabular}{crrrrrrrrr}
    \hline
    \multicolumn{10}{|c|}{\textbf{EUCLID survey}} \\
    \hline
    \multicolumn{10}{|c|}{\textbf{$k_{\max}=0.2\,h\,\mathrm{Mpc}^{-1}$}} \\
    \multicolumn{10}{|c|}{$\sigma(\beta)=2.2\times 10^{-1}$ ($\frac{\sigma(\beta)}{\beta_{\rm fid}}=1.47$)\hspace{1em}$\sigma(\Omega_{m0})=5.32\times 10^{-2}$ ($\frac{\sigma(\Omega_{m0})}{\Omega_{m0,\rm fid}}=0.172$)} \\
    \multicolumn{10}{|c|}{$\sigma(\ln P_0)$: min 0.0989, median 0.154, max 0.189} \\
    \hline
    $z$ & $\ln b_1$ & $b_2$ & $b_{G_2}$ & $b_{\Gamma_3}$ & $c_0$ & $c_2$ & $\tilde c$ & $\ln\sigma_f$ & $P_{\rm shot}$ \\
    \hline
    1 & 0.108 & 3.65 & 1.58 & 5.76 & 83.1 & 26.2 & 335 & 0.211 & 0.38 \\
    1.2 & 0.119 & 4.58 & 1.82 & 6.91 & 88.6 & 29.6 & 313 & 0.207 & 0.311 \\
    1.4 & 0.129 & 5.63 & 2.21 & 8.29 & 93.1 & 33.3 & 317 & 0.217 & 0.24 \\
    1.65 & 0.142 & 7.06 & 2.77 & 10.3 & 97.4 & 36.7 & 323 & 0.228 & 0.147 \\
    \hline
    \multicolumn{10}{|c|}{\textbf{$k_{\max}=0.25\,h\,\mathrm{Mpc}^{-1}$}} \\
    \multicolumn{10}{|c|}{$\sigma(\beta)=8.6\times 10^{-2}$ ($\frac{\sigma(\beta)}{\beta_{\rm fid}}=0.573$)\hspace{1em}$\sigma(\Omega_{m0})=2.12\times 10^{-2}$ ($\frac{\sigma(\Omega_{m0})}{\Omega_{m0,\rm fid}}=0.068$)} \\
    \multicolumn{10}{|c|}{$\sigma(\ln P_0)$: min 0.0515, median 0.119, max 0.142} \\
    \hline
    $z$ & $\ln b_1$ & $b_2$ & $b_{G_2}$ & $b_{\Gamma_3}$ & $c_0$ & $c_2$ & $\tilde c$ & $\ln\sigma_f$ & $P_{\rm shot}$ \\
    \hline
    1 & 0.05 & 1.52 & 1.16 & 3.52 & 40.4 & 13.4 & 189 & 0.127 & 0.141 \\
    1.2 & 0.0531 & 1.96 & 1.42 & 4.46 & 43.5 & 14 & 181 & 0.133 & 0.104 \\
    1.4 & 0.0562 & 2.46 & 1.77 & 5.54 & 46.3 & 15.2 & 185 & 0.143 & 0.0829 \\
    1.65 & 0.0598 & 3.16 & 2.26 & 7.03 & 49 & 16.6 & 185 & 0.148 & 0.0541 \\
    \hline
    \multicolumn{10}{|c|}{\textbf{$k_{\max}=0.3\,h\,\mathrm{Mpc}^{-1}$}} \\
    \multicolumn{10}{|c|}{$\sigma(\beta)=6.02\times 10^{-2}$ ($\frac{\sigma(\beta)}{\beta_{\rm fid}}=0.401$)\hspace{1em}$\sigma(\Omega_{m0})=1.67\times 10^{-2}$ ($\frac{\sigma(\Omega_{m0})}{\Omega_{m0,\rm fid}}=0.054$)} \\
    \multicolumn{10}{|c|}{$\sigma(\ln P_0)$: min 0.04, median 0.0711, max 0.0828} \\
    \hline
    $z$ & $\ln b_1$ & $b_2$ & $b_{G_2}$ & $b_{\Gamma_3}$ & $c_0$ & $c_2$ & $\tilde c$ & $\ln\sigma_f$ & $P_{\rm shot}$ \\
    \hline
    1 & 0.0397 & 1.24 & 0.919 & 2.87 & 31.7 & 7.16 & 114 & 0.0773 & 0.0623 \\
    1.2 & 0.0424 & 1.62 & 1.16 & 3.67 & 34.7 & 8.4 & 114 & 0.083 & 0.0521 \\
    1.4 & 0.0448 & 2.05 & 1.46 & 4.57 & 37.3 & 9.71 & 119 & 0.0912 & 0.0444 \\
    1.65 & 0.0476 & 2.63 & 1.89 & 5.82 & 39.7 & 10.8 & 121 & 0.0962 & 0.0326 \\
    \hline
  \end{tabular}
  }
\end{table*}

\section{\label{sec:C}Conclusion}

In this work, we forecast constraints on the parameters of the CDE model using the one-loop galaxy power spectrum of DESI and Euclid, and explore ways to improve them.  Extending the analysis to mildly nonlinear scales provides additional information and substantially tightens the constraints. For example, for $\beta_{\rm fid}=0.15$ and $\nu=0.1$, increasing $k_{\max}$ from $0.1$ to $0.2\,h\,{\rm Mpc}^{-1}$ reduces $\sigma(\beta)$ from $0.692$ to $0.139$, giving a gain of $4.96$. Combining surveys also improves the constraints by extending the redshift coverage and increasing the galaxy sample and survey volume. Using all seven DESI bins together with the two highest-redshift Euclid bins improves the coupling constraint by factors of $1.60$ and $1.58$ relative to DESI and Euclid alone, respectively.
We further investigate how a prior on $\Omega_{m0}$ affects its degeneracy with $\beta$, and how a smoothing prior between neighbouring power-spectrum bands changes the constraints. A prior that is small compared with the original band-power Fisher matrix can still have a sizable effect when it acts on a weakly constrained, degenerate parameter combination. We find however that while the $\Omega_{m0}$ prior has a strong effect on $\beta$, the cross-band prior on $P_0(k)$ reduces the bounds on the coupling by less than a few percent. 

The main goal of this paper was to assess the distinguishability of CDE from the uncoupled model, since recent analyses favour small coupling values~\cite{gómezvalent2026, Chakraborty:2025syu}. We find
that the constraint on $\beta$ weakens as its fiducial value approaches zero. At $\nu=0.1$ and our reference $k_{\max}=0.2\,h\,{\rm Mpc}^{-1}$, the lowest sampled value distinguishable from zero at $1\sigma$ is $\beta_{\rm fid}=0.15$. This threshold can be lowered by weakening the $\beta$--$\Omega_{m0}$ degeneracy: an illustrative Gaussian prior with $\sigma_{\Omega,\rm ext}=0.0057$ allows $\beta_{\rm fid}=0.08$ to be distinguished from zero at approximately $1\sigma$. 

In the most optimistic case, i.e. assuming the joint survey, $k_{\rm max}=0.3h/$Mpc, and the Gaussian prior on $\Omega_{m0}$, we find that $\beta=0.05$ can be distinguished from zero to 1$\sigma$. Our threshold values for $\beta$ are substantially larger than the best fit currently found from combined early- and late-time cosmological data \cite{Chakraborty:2025syu,gómezvalent2026}. These best fit, however, have been obtained assuming a constant coupling from decoupling to today, while we only examined late-time data. The possibility of detecting the coupling with late-time data seems therefore to require models with time-varying couplings, so that one can have a small $\beta$ in the past and a relatively larger one today. Constructing and motivating such models is an interesting challenge for future work. 

\appendix

\section{\label{app:prior-gains} Appendix }
We study how a  prior on the \(P_0(k_i)\) bands affects the marginalized errors on \(\beta\) and the band powers.  We first focus on \(\beta\). We write the global parameters \(\boldsymbol{g}=(\beta,\Omega_{m0})\) explicitly in the Fisher matrix of Eq.~\eqref{eq:mat1}, while leaving the redshift-bin structure of the nuisance parameters  implicit. Adding the prior matrix \(\Pi\) to the linear-spectrum block \(F_{pp}\) gives

\begin{equation}
F
=
\begin{pmatrix}
F_{\beta\beta}
&
F_{\beta\Omega} & F_{\beta p}
&
F_{\beta n}
\\
F_{\Omega\beta} & F_{\Omega\Omega} & F_{\Omega p} & F_{\Omega n} \\ F_{p\beta} & F_{p\Omega}
&
F_{pp}+ \Pi
&
F_{pn}
\\
F_{n\beta} & F_{n\Omega}
&
F_{np}
&
F_{nn}
\end{pmatrix}.
\label{eq:complete_fisher_prior}
\end{equation}

To separate the $F_{\beta\beta}$  block from the rest, we  define 
\begin{equation}
\boldsymbol{q}
=
\left(\Omega_{m0},\boldsymbol{p},\boldsymbol{n}\right),\qquad\Pi_q\equiv\begin{pmatrix}0 & 0 & 0\\0 & \Pi & 0\\0 & 0 & 0\end{pmatrix},
\end{equation}
where $\boldsymbol{q}$ denotes the parameter subspace excluding $\beta$,  $\Pi_q$ embeds the band-power prior within this subspace, and $F_{qq}$ denotes the corresponding Fisher-matrix block. The Fisher matrix can then be written compactly as
\begin{equation}
F
=
\begin{pmatrix}
F_{\beta\beta} & F_{\beta q}\\
F_{q\beta} & F_{qq}+\Pi_q
\end{pmatrix}.
\end{equation}
After marginalizing over $\Omega_{m0}$, all band-power parameters, and all nuisance parameters, the remaining information on $\beta$ is given by the Schur complement,
\begin{equation}
\sigma_\beta^{-2}
=
F_{\beta\beta}
-
F_{\beta q}\left[F_{qq}+\Pi_q\right]^{-1}F_{q\beta},
\label{eq:beta_marginal_information}
\end{equation}
The second term in Eq.~\eqref{eq:beta_marginal_information} is the information lost through degeneracies between $\beta$, $\Omega_{m0}$, the  $P_0(k_i)$ bands , and the nuisance parameters. The  prior enters explicitly through $F_{qq}+\Pi_q$. Increasing $\Pi_{\rho}$ strengthens the band-power sector and reduces the part of this degeneracy associated with spectral-shape variations, thereby improving the marginalized constraint on $\beta$. 
For an infinitesimal  prior:
\begin{equation}
\Pi_q=\varepsilon P_q,\qquad \varepsilon\to0.
\label{eq:app_small_prior}
\end{equation}
Expanding the inverse factor to first order gives:
\begin{equation}
(F_{qq}+\varepsilon P_q)^{-1}
=
F_{qq}^{-1}
-\varepsilon F_{qq}^{-1}P_qF_{qq}^{-1}
\label{eq:app_inverse_expansion}
\end{equation}
Substituting this expansion into Eq.~\eqref{eq:beta_marginal_information} gives:
\begin{equation}
\sigma_\beta^{-2}(\varepsilon)
=
\underbrace{
F_{\beta\beta}-F_{\beta q}F_{qq}^{-1}F_{q\beta}
}_{\displaystyle \sigma_{\beta,0}^{-2}}
+
\varepsilon
F_{\beta q}F_{qq}^{-1}P_qF_{qq}^{-1}F_{q\beta}
\label{eq:app_beta_information_expansion}
\end{equation}
Here, $\sigma_{\beta,0}^{2}$ is marginalized constraint without prior and $\sigma_\beta^{-2}(\varepsilon)$ is the one with prior. Define
\begin{equation}
v=F_{qq}^{-1}F_{q\beta}.
\label{eq:app_degeneracy_vector}
\end{equation}
The result then becomes
\begin{equation}
\sigma_\beta^{-2}(\varepsilon)
=
\sigma_{\beta,0}^{-2}
+\varepsilon\,\boldsymbol v^{\rm T}P_q\boldsymbol v
\label{eq:app_beta_information_compact}
\end{equation}

The vector $\boldsymbol v$ identifies the combination of the other parameters that is degenerate with \(\beta\). The prior affects $\beta$ according to its action along this combination. The corresponding gain is
\begin{equation}
G_\beta
\equiv
\frac{\sigma_{\beta,0}}{\sigma_\beta(\varepsilon)}
=
1+
\frac{\varepsilon}{2}\,
\sigma_{\beta,0}^{2}
\boldsymbol v^{\rm T}P_q\boldsymbol v.
\label{eq:app_beta_gain}
\end{equation}
A small prior can have an amplified effect when the original marginalized uncertainty $\sigma_{\beta,0}^{2}$ is large or \(v\) amplifies a weakly constrained direction.  For example, if $F_{q\beta}$ is zero (when there is no correlation between $\beta$ and other parameters),  there is no gain. Therefore, we can say that parameters that strongly correlate with band power and have weak constraints show greater improvement.

Having derived the gain for $\beta$ in Eq.~\eqref{eq:app_beta_gain}, we now obtain the corresponding relation for the band powers. We regroup the parameter space as
\begin{equation}
\boldsymbol p=\bigl(\ln P_0(k_1),\ldots,\ln P_0(k_{N_k})\bigr),
\qquad
\boldsymbol r=(\beta,\Omega_{m0},\boldsymbol n).
\label{eq:app_parameter_partition}
\end{equation}
We now retain $\boldsymbol p$ and marginalize over all remaining parameters, including $\beta$. The Fisher matrix before and after adding the band-power prior is
\begin{equation}
F=
\begin{pmatrix}
F_{pp} & F_{pr}\\
F_{rp} & F_{rr}
\end{pmatrix},
\qquad
F^{\rm prior}=
\begin{pmatrix}
F_{pp}+\Pi & F_{pr}\\
F_{rp} & F_{rr}
\end{pmatrix}.
\label{eq:app_band_fisher}
\end{equation}
Applying the Schur complement gives the marginalized band-power  information,
\begin{equation}
S=F_{pp}-F_{pr}F_{rr}^{-1}F_{rp},
\label{eq:app_band_schur}
\end{equation}
and, with the prior,
\begin{equation}
S_{\rm prior}
=(F_{pp}+\Pi)-F_{pr}F_{rr}^{-1}F_{rp}
=S+\Pi.
\label{eq:app_band_schur_prior}
\end{equation}
 Inverting these matrices gives the band covariances,
\begin{equation}
C_{pp}^{\,0}=S^{-1},
\qquad
C_{pp}^{\,\rm prior}=(S+\Pi)^{-1}.
\label{eq:app_band_covariances}
\end{equation}
Their diagonal entries give the marginalized errors for band $i$,
\begin{equation}
\sigma_{i,0}=\sqrt{(S^{-1})_{ii}},
\qquad
\sigma_{i,\rm prior}
=\sqrt{\bigl[(S+\Pi)^{-1}\bigr]_{ii}},
\label{eq:app_band_errors}
\end{equation}
and their ratio defines the gain,
\begin{equation}
G_i(\rho,k_{\rm corr})
=\frac{\sigma_{i,0}}{\sigma_{i,\rm prior}}
=\sqrt{
\frac{(S^{-1})_{ii}}
{\left[\left(S+\Pi(\rho,k_{\rm corr})\right)^{-1}\right]_{ii}}
}.
\label{eq:app_band_gain}
\end{equation}

The term $F_{pr}F_{rr}^{-1}F_{rp}$ in Eq.~\eqref{eq:app_band_schur} represents information lost through degeneracies with the remaining parameters. If a combination of band powers is strongly degenerate with some other parameters, for example $c_0$, the marginalized information \(S\) can be much smaller than the  \(F_{pp}\). A prior $\Pi(\rho,k_{\rm corr})$ that is small compared with $F_{pp}$ can therefore be substantial relative to this remaining information S. The relevant comparison is thus with $S$, not merely with $F_{pp}$. Therefore, the prior has a larger effect on bands that are strongly correlated with other parameters. Fixing these parameters substantially reduces this effect, as shown in Figure~\ref{fig:band_gain_fixed}.

\bibliography{apssamp}
\end{document}